\documentclass[12pt]{article}

\def\lromn#1{\uppercase\expandafter{\romannumeral#1}}

\def\blist{\begin{list}{\setlength{\rightmargin}{\leftmargin}}}
\def\elist{\end{list}}
\begin{document}

\begin{flushright}
TU/00/592 \\
KEK-TH-697\\
\end{flushright}

\begin{center}
\begin{large}

\textbf{
Dynamics of barrier penetration in thermal medium: exact result
for inverted harmonic oscillator
}

\end{large}

\vspace{36pt}

\begin{large}
Sh. Matsumoto$^{1}$ and M. Yoshimura$^{2}$

$^{1}$ 
Theory Group, KEK\\
Oho 1-1 Tsukuba Ibaraki 305-0801 Japan

$^{2}$ 
Department of Physics, Tohoku University\\
Sendai 980-8578 Japan\\
\end{large}

\vspace{4cm}

{\bf ABSTRACT}
\end{center}

\vspace{1cm}
Time evolution of quantum tunneling is studied when
the tunneling system is immersed in thermal medium.
We analyze in detail the behavior of the system after
integrating out the environment.
Exact result for the inverted harmonic oscillator of the
tunneling potential is derived and the barrier penetration
factor is explicitly worked out as a function of time.
Quantum mechanical formula without environment
is modifed both by the potential
renormalization effect and by a dynamical factor which
may appreciably differ from the previously obtained one
in the time range of 1/(curvature at the top of potential
barrier).

\newpage

\lromn 1
\hspace{0.2cm}
{\bf Introduction}

\vspace{0.5cm} 
Tunneling phenomena in thermal medium are of both theoretical
and practical interest in many areas of physics.
For instance, in cosmology there are a variety of tunneling phenomena that may
have occured in the evolution of our universe.
If the phase transition of the electroweak gauge
symmetry is of the first order as assumed in the electroweak
scenario of baryogenesis \cite{ew-bgeneration-review}, 
then the tunneling from a metastable
state to the true ground state of the electroweak theory must
go through either quantum effect or thermally activated barrier
crossing.
Another possibility of the first order phase transition in cosmology is the
quark-hadron phase transition.
We should neither fail to mention the classic example of tunneling
phenomenon that takes place in the central core of stars:
thermonuclear reaction \cite{clayton}.

A common problem to all these is the presence of environment:
the tunneling we are interested in does not take place for the system
in isolation.
Under this circumstance we are very much interested in how substantially
quantum mechanical formula for the tunneling rate is modified
by dissipative interaction with surrounding medium.
Despite of this obvious interest, many past works in cosmology
and astropysics have relied 
on simple methods to deal with the tunneling problem, 
either by using the bounce solution in the Euclidean approach
\cite{langer}, \cite{callen-coleman}, \cite{affleck}
or by exploiting some variant of the classical Vant Hoff-Arrhenius law
\cite{kramers}, \cite{q-tunneling review}.

On the other hand, since the pioneering work \cite{caldeira-leggett 83}
on quantum dissipation to tunneling phenomena at zero temperature,
there have appeared many extended works 
in condensed matter or statistical physics community
(a partial list of these works is given by
\cite{larkin-ovchinnikov}, \cite{grabert et al-87}, \cite{waxman-leggett}).
Intensive theoretical activities in this field are presumably related to
experimental possibility of observing the macroscopic quantum tunneling
in various areas of condensed matter physics.
The problem is however not simple and only a limited class of problems
have been addressed.  Thus even in idealized models
one often assumes that the entire system is in thermal equilibrium
and attempts to derive quantities of limited value such as the decay rate of
the metastable state by an extention of the bounce analysis.
In some of these works \cite{larkin-ovchinnikov}, \cite{grabert et al-87}, 
the key quantity is the imaginary
part of the free energy, which may be interpreted as the decay rate.
However, since equilibrium value of the free energy of the entire system is 
necessarily real,
one must extract the imaginary part by projecting to the initial metastable
state. In some discussions in the literature it is not clear how this
projection is performed on rigorous grounds, 
although some of its physical consequences are reasonable.
There is indeed some criticism againt this type of approach
\cite{waxman-leggett}, \cite{zweger}.
It seems that more fundamental microscopic approach to deal with the tunneling
in dissipative medium is needed.

With this background in mind, 
our aim in the present work is to clarify dynamical aspects of the tunneling
in medium:
how the barrier penetration basic to quantum tunneling proceeds with real time.
The key idea is separation of a small system from a larger environment,
and we would like to determine the reduced density matrix for the small system
by integrating out environment variables \cite{feynman-vernon}.
This makes it easy to compute the penetrating flux factor for
an energy eigenstate of the small subsystem.
Our approach does not use the Euclidean technique, which in our opinion
obscures dynamics of the time evolution.
We neither assume that the tunneling system is in thermal
equilibrium with environment, although we can discuss this case
using our fundamental formula.
Moreover, we are able to deal with both the quantum and the thermally
activated regions in a unified way.

The model we use to extract exact results for the barrier penetration is 
the inverted harmonic oscillator (IVHO). 
Since the form of the potential we use for exact results is very special, 
we cannot discuss the tunneling for general cases with full confidence.
Nevertheless we believe that the method employed in the present
work, especially the integral transform of the Wigner function,
should be useful to derive approximate, yet valuable results for
general potential in the semiclassical approach.
We hope to return to a general tunneling potential in our
subsequent work.

We take for the environment an infinitely many harmonic oscillators
of arbitrary spectrum.
This is a standard one studied by many people in the field.
The system of a normal harmonic oscillaor coupled to this
environment is analytically solvable, as discussed
in \cite{jmy-96-1}, \cite{jmy-96-2}.
Our barrier penetration model corresponds to a case of imaginary frequency.
In the present work we shall employ and extend some techniques we developed
in the case of normal oscillator.

The main result of the present work is summarized by a formula
for the barrier penetration;
\begin{eqnarray}
&&
f(t) \,|T(E)|^{2}  \,,
\hspace{0.5cm} 
f(t) = \frac{p_{0}(t)}{\omega _{B}\,q_{0}(t)} =
\frac{1}{\omega _{B}}\,\frac{d}{dt}\,\ln |q_{0}(t)|
\,.
\label{fundamental formula 0} 
\end{eqnarray}
This formula is applied to an eigenstate of energy $E$
taken for the initial IVHO subsystem.
The well known quantum mechanical penetration formula
is modified by the environment effect in two ways;
first, via the change of the original curvature
$\omega _{0}$ (bare parameter)
to the renormalized, pole curvature $\omega _{B}$;
\begin{equation}
|T(E)|^{2} = \frac{1}{1 + e^{2\pi|E|/\omega _{0}}} 
\longrightarrow
\frac{1}{1 + e^{2\pi |E|/\omega _{B}}} 
\,,
\end{equation}
with $E$ the energy measured from the top of the potential barrier.
This effect is essentially similar to, but numerically different from,
the curvature renormalization effect much emphasized by Caldeira and Leggett
\cite{caldeira-leggett 83}.
In their work two cases with and without the friction term, but
both including the curvature renormalization given by $\omega _{R}$
(which is larger than $\omega _{B}$) are compared. 
The result of Caldeira and Leggett is understood when one writes
$\omega _{B}$ in terms of $\omega _{R}$.

The second environment effect is 
the time dependent factor $f(t)$ which is a ratio
of the momentum $p_{0}(t) = \dot{q}_{0}(t)$ to the rescaled
coordinate trajectory $q_{0}(t)$.
This trajectory function $q_{0}(t)$ obyes the homogeneous Langevin
equation, eq.(\ref{langevin eq}), 
under thermal environment, being characterized by
the initial energy corresponding to the top of the potential barrier.
When the dynamical function $f(t) = 1$, 
the IVHO subsystem has the energy of the barrier top,
and its deviation from unity is a measure of energy flow from
the environment.
When $f(t) > 1$, namely $|p_{0}(t)| > \omega _{B}\,|q_{0}(t)|$,
the IVHO system is excited by an energy inflow from the environment.
On the other hand, when $f(t) < 1$, the system is deexcited by an
energy outflow.
The main new factor $f(t)$ deviates from unity within time
range of order $1/\omega _{B}$, and both for $t \ll 1/\omega _{B}$
and for $t \gg 1/\omega _{B}$ the effect is small;
$f(t) \approx 1$.
We find interesting examples in which this dynamical function
exceeds unity, thus implying enhanced penetration, albeit for
a short time of period of order $1/\omega _{B}$.

The rest of this paper is organized as follows.
In Section \lromn 2 we explain how we model the environment and
its interaction with a quantum system, and introduce the influence
functional. Quantum Langevin equation is also briefly touched upon.
In Section \lromn 3 we work out exact consequences for the inverted
harmonic oscillator, and give the barrier penetration factor
taking an energy eigenstate for the initial state.
Our general result includes an integral transform of the Wigner
function from the initial to the final one,
as explained in Appendix A, .
The fundamental formula (\ref{fundamental formula 0}) is derived
along with an explicit form of $q_{0}(t)/q_{0}(0)$,
eq.(\ref{zero energy motion}).
The  Ohmic or the local approximation 
(the inherently non-local term of
environment interaction in the full Langevin equation being replaced 
by a few expansion terms)
is shown to lead to some anomalous behavior of the dynamical factor.
In Section \lromn 4 some applications to the mixed initial
state are discussed using the exact
result of the inverted harmonic oscillator.
Our understanding of the result of ref.\cite{caldeira-leggett 83}
is also made in terms of the diagonalized curvature parameter
$\omega _{B}$.
Three appendices explain technical points somewhat off the
main stream of arguments in the text.
Appendix A gives the interesting form of the integral transform of 
the  Wigner function, while Appendix B gives the differential form
of the Fokker-Planck equation, derived both for the harmonic model
of environment.
Appendix C explains parameters necessary for our numerical
computation of the dynamical function.

\vspace{1cm}
\lromn 2 \hspace{0.2cm}
{\bf Model of environment and influence functional method
}

\vspace{0.5cm} 
We expect that 
the behavior of a small system immersed in thermal environment
is insensitive to detailed modeling of the environment and
the form of its interaction to the system. 
Only global quantities such as the environment temperature,
the friction, the threshold of the environment spectrum,
are expected to be important.
Since the pioneering work of Feynman-Vernon \cite{feynman-vernon}  and 
Caldeira-Leggett \cite{caldeira-leggett 83} the standard model uses
an infinite set of harmonic oscillators (its coordinate variable
denoted by $Q(\omega )$)
for the environment and a bilinear interaction with the small system
(denoted by $q$);
\begin{eqnarray}
&& \hspace*{-0.5cm}
L_{Q} = \frac{1}{2}\, \int_{\omega _{c}}^{\infty }\,d\omega \,
\left( \,\dot{Q}^{2}(\omega )- \omega ^{2}\,Q^{2}(\omega )\,\right) \,,
\hspace{0.5cm} 
L_{{\rm int}} = -\, q\,\int_{\omega _{c}}^{\infty }\,d\omega \,
c(\omega )Q(\omega ) \,. 
\label{harmonic environment}
\end{eqnarray}
We assume existence of a threshold $\omega _{c} > 0$.
The coupling strength to the environment is specified by
$c(\omega )$.
In the present section we do not assume any special form for
the potential of the $q-$system $V(q)$.

It is now appropriate 
to explain the influence functional method \cite{feynman-vernon}.
The influence functional denoted by ${\cal F}[q(\tau ) \,, q'(\tau )]$ 
results after integration of the environment variable $Q(\omega )$ 
when one computes the density matrix of the entire system.
Since the density matrix is a product of the transition
amplitude and its conjugate, the path integral formula
resulting from the environment integration has a functional
dependence both on the system path $q(\tau )$ and its conjugate path
$q'(\tau )$.
We thus define, when an initial environment is in a mixed state 
given by a density matrix
\( \:
\rho_{i} (Q_{i}\,, Q'_{i}),
\: \)
the influence functional
\begin{eqnarray}
&& \hspace*{1cm}
{\cal F}[\,q(\tau )\,, q'(\tau )\,] \equiv
\int\,{\cal D}Q(\tau )\,\int\,{\cal D}Q'(\tau )\,\int\,dQ_{i}\,
\int\,dQ'_{i}\,
\nonumber \\
&& \hspace*{-1cm}
\int\,dQ_{f}\,\int\,dQ'_{f}
\,\delta (Q_{f} - Q'_{f})
\cdot K\left( \,q(\tau )\,,Q(\tau ) \,\right)\,
K^{*}\left( \,q'(\tau )\,,Q'(\tau ) \,\right)\,\rho_{i} (Q_{i}\,, Q'_{i}) \,,
\label{def of influence functional}
\\ && \hspace*{1cm} 
K\left( \,q(\tau )\,,Q(\tau ) \,\right) =
\exp \left( \,iS_{Q}[Q] + iS_{{\rm int}}[q \,, Q]\,\right) \,, 
\\ && \hspace*{1cm} 
S_{Q}[Q] + S_{{\rm int}}[q \,, Q] = \int_{0}^{t}\,d\tau \,
\left( \,L_{Q}[Q] + L _{{\rm int}}[q,\,, Q]\,\right) \,,
\\ && \hspace*{2cm} 
\rho _{i}(Q_{i}\,, Q_{i}\,') =
\sum_{n}\,w_{n}\,\psi _{n}^{*}(Q_{i}\,')\psi _{n}(Q_{i}) 
\,, 
\end{eqnarray}
where $w_{n}$ is the probability of finding a pure state $n$
in the initial environment.
The fact that there is a delta function $\delta (Q_{f} - Q'_{f})$ 
for the environment variable at the final time $t$
indicates that one does not observe the environment part at $t$.
Throughout this discussion we consider a definite time interval,
$0 < \tau < t$.

For the thermal ensemble of a $Q$ harmonic oscillator of frequency $\omega $
($\beta = 1/T$ being the inverse temperature),
the density matrix is
\begin{eqnarray}
&&
\hspace*{2cm}
\rho _{\beta }(Q \,, Q') = 
\nonumber \\ && \hspace*{-1cm}
\left( \frac{\omega }{\pi \,\coth (\beta \omega 
/2)}\right)^{1/2}\,
\exp \left[ \,-\,\frac{\omega }{2 \sinh (\beta \omega )}
\,\left( \,(Q^{2} + Q'\,^{2})\,\cosh (\beta \omega ) 
- 2 Q Q'\,\right)\,\right]
\,, 
\label{thermal density matrix} 
\end{eqnarray}
and one can explicitly perform the $Q(\omega )$ path integration
in eq.(\ref{def of influence functional}),
since the $Q(\omega )$ integration is Gaussian.
The result is a nonlocal action;
\begin{eqnarray}
&& \hspace*{-1cm}
{\cal F}[\,q(\tau )\,, q'(\tau )\,] =
\exp [\,-\,\int_{0}^{t }\,d\tau \,\int_{0}^{\tau }\,ds\,
\left( \,\xi (\tau )\alpha_{R}(\tau - s)\xi (s) + i\xi (\tau )
\alpha _{I}(\tau - s)X(s)\,\right)\,] \,, 
\nonumber \\ &&
\\ &&
\hspace*{1cm} {\rm with} \hspace{0.2cm}
\xi (\tau ) = q(\tau ) - q'(\tau ) \,, \hspace{0.5cm} 
X(\tau ) = q(\tau ) + q'(\tau ) \,,
\\ && \hspace*{1cm} 
\alpha (\tau ) \equiv \alpha _{R}(\tau ) + i\alpha _{I}(\tau )
= \frac{1}{2\pi }\,\int_{-\infty }^{\infty }\,d\omega \,
\tilde{\alpha }(\omega )\,e^{-i\omega \tau } \,, 
\\ &&
\tilde{\alpha }(\omega ) =
i\,\int_{\omega _{c}}^{\infty }\,d\omega '\,c^{2}(\omega ')\,
\left( \,\frac{1}{\omega ^{2} - \omega '\,^{2} + i0^{+}}
- \frac{2\pi i}{e^{\beta \omega '} - 1}\,
\delta (\omega'\, ^{2} - \omega^{2})\,\right) \,.
\end{eqnarray}
Both of $\alpha _{R}$ and $\alpha _{I}$ are real functions.
A complex combination of these, the kernel function $\alpha (\tau )$
that appears in the exponent of the influence functional
is the real-time thermal Green's function
for a collection of harmonic oscillators $Q(\omega )$,
which makes clear a relation to thermal field theory.

The reduced density matrix $\rho ^{(R)}$
for the $q$ system is defined as follows.
For simplicity we take for the initial system  a pure quantum state
given by a wave function $\psi (q_{i})$;
\begin{eqnarray}
&& \hspace*{2cm}
\rho ^{(R)}(q_{f} \,, q'_{f}) =
\nonumber \\ &&
\int\,{\cal D}q(\tau )\,\int\,{\cal D}q'(\tau )\,
\int\,dq_{i}\,\int\,dq'_{i}\;
\psi ^{*}(q'_{i})\,{\cal F}[\,q(\tau )\,, q'(\tau )\,]\,
e^{iS_{q}[q] - iS_{q}[q']}\,\psi (q_{i})
\,. 
\end{eqnarray}
Here $S_{q}[q]$ is the action for the $q-$system.
This density matrix $\rho ^{(R)}$
may be used to compute physical quantities of one's interest.

It is sometimes convenient to introduce the Wigner function by
\begin{eqnarray}
&&
f_{W}^{(R)}(x \,, p) = \int_{-\infty }^{\infty }\,
\frac{d\xi }{\sqrt{2\pi }}\,\rho ^{(R)}\left( x + \frac{\xi }{2}
\;, x - \frac{\xi }{2} \right)\,e^{-ip\xi }  \,.
\label{def of wigner}
\end{eqnarray}
We shall later mention the master equation for $f_{W}^{(R)}$.
Here we quote for comparison a master equation for the Wigner
function when the entire system is in a pure quantum state;
\begin{eqnarray}
&&
\frac{\;\partial f_{W}}{\partial t} = 
-\,p\,\frac{\partial f_{W}}{\partial x} + 
\frac{1}{i\hbar }\left( \,
V(x + \frac{i\hbar }{2}\frac{\partial }{\partial p})
- V(x - \frac{i\hbar }{2}\frac{\partial }{\partial p})
\,\right)\,f_{W} 
\,.
\end{eqnarray}
It takes a form of infinite dimensional differential equation.

The master equation is simplified in the semiclassical limit
of $\hbar \rightarrow 0$ to
\begin{eqnarray}
&&
\frac{\;\partial f_{W}}{\partial t} = 
-\,p\,\frac{\partial f_{W}}{\partial x} + 
\frac{\partial V}{\partial x}\,\frac{\partial f_{W}}{\partial p}
\,. \label{semiclassical liouville} 
\end{eqnarray}
A great virtue of the Wigner function is that this semiclassical equation
coincides with the Liouville equation for the distribution
function  of a classical statistical system in the phase space
$(x \,, p)$.
It is thus easy to write down the simiclassical solution in
the form of an integral transform,
\begin{equation}
f_{W}^{}(x\,, p) = 
\,\,\int_{-\infty }^{\infty }\,dx_{i}\,\int_{-\infty }^{\infty }\,dp_{i}
\,f_{W}^{(i)}(x_{i}\,, p_{i})
\,\delta \left( x - \tilde{x}_{{\rm cl}}(t)\right)\,
\delta \left( p - \tilde{p}_{{\rm cl}}(t) \right) \,, 
\end{equation}
where the classical deterministic flow,
\( \:
(x_{i} \,, p_{i}) \rightarrow 
(\tilde{x}_{{\rm cl}} \,, \tilde{p}_{{\rm cl}}) \,, 
\: \)
is defined by the classical mapping satisfying
\begin{equation}
\tilde{p}_{{\rm cl}} = \frac{d\tilde{x}_{{\rm cl}}}{dt} \,, \hspace{0.5cm} 
\frac{d\tilde{p}_{{\rm cl}}}{dt} = -\,
\left( \frac{\partial V}{\partial x}\right)_{x = \tilde{x}_{{\rm cl}}}
\,.
\end{equation}

Although it is not our main tool of analysis, it might be of some use
to recall the quantum Langevin equation for the model of 
eq.(\ref{harmonic environment}) \cite{ford-lewis-oconnell}. 
By eliminating the environment variable
$Q(\omega \,, t)$ one derives the operator equation for the $q$
variable;
\begin{eqnarray}
&&
\frac{d^{2}q}{dt^{2}} + \frac{\partial V}{\partial q} +
2\,\int_{0}^{t}\,ds\,\alpha _{I}(t - s)q(s) = F_{Q}(t) 
\,,
\\ &&
F_{Q}(t) = -\,
\int_{\omega _{c}}^{\infty }\,d\omega \,c(\omega )\,
\left( \,Q(\omega \,, 0) \cos (\omega t) +
\frac{P(\omega \,, 0)}{\omega } \sin (\omega t)\,\right)
\,, 
\\ &&
\langle F_{Q}(\tau )F_{Q}(s)^{{\rm sym}} \rangle_{{\rm env}}
= \int_{\omega _{c}}^{\infty }\,d\omega \,r(\omega )
\cos \omega (\tau - s)\,\coth (\frac{\beta \omega }{2}) \,,
\end{eqnarray}
where $r(\omega) = c^{2}(\omega)/(2\omega)$
and $\langle \; \rangle_{{\rm env}}$ is the thermal average over
the environment variables.
Thus, the kernel function $\alpha _{I}(\tau )$ describes
a nonlocal action from the environment, while $F_{Q}(t)$
is a random force from the environment.
The local approximation to $\alpha _{I}$,
taking the form of
\( \:
\alpha _{I}(\tau ) = \delta \omega ^{2}\delta (\tau )
+ \eta\, \delta '(\tau ) \,, 
\: \)
gives 
\begin{equation}
\frac{d^{2}q}{dt^{2}} + \frac{\partial V}{\partial q} +
\delta \omega ^{2}\,q + \eta \frac{dq}{dt} = F_{Q}(t) \,.
\label{langevin eq in ohmic approx} 
\end{equation}
The parameter $\delta \omega ^{2}$ in this case is the frequency
shift due to the presence of the environment, and
$\eta $ is the Ohmic friction.
We call this approximation the Ohmic approximation.
Perhaps more suitably, it might better be called the local
approximation.

On the other hand, the real part of the kernel 
function $\alpha _{R}$ describes 
fluctuation, and it is related to the dissipation $\alpha _{I}$ by
the fluctuation-dissipation theorem;
for their Fourier components
\begin{equation}
\tilde{\alpha }_{I}(\omega ) = \frac{1}{\pi }\,
{\cal P}\,\int_{\omega _{c}}^{\infty }\,
d\omega '\,\frac{2\omega ' \tilde{\alpha }_{R}(\omega ')\,
\tanh (\beta \omega '/2)}{\omega ^{2} - \omega '\,^{2}}
\,,
\end{equation}
where ${\cal P}$ denotes the principal part of integration.

\vspace{1cm}
\lromn 3 \hspace{0.2cm}
{\bf Exact results for inverted harmonic oscillator
}

\vspace{0.5cm} 
\lromn 3A \hspace{0.2cm}
{\bf Formalism}

\vspace{0.5cm} 
We specialize the system dynamics of barrier penetration to
that of the inverted harmonic oscillator (IVHO) given by
the Lagrangian density,
\begin{eqnarray}
L_{q} 
= \frac{1}{2}\, \dot{q}^{2} - V_{q}(q) \,, \hspace{0.5cm} 
V_{q}(q) = -\,\frac{1}{2}\, \omega_{0} ^{2}\,q^{2}
\,, \hspace{0.5cm} \omega _{0}^{2} > 0 \,.
\end{eqnarray}
There are similarities to the case of a normal oscillator of
$\omega_0^2 < 0$, and
we can take over some of the results derived for the unstable 
($|\omega _{0}|^{2} > \omega _{c}^{2}$) \cite{jmy-96-1}, \cite{jmy-96-2}
or the stable ($0 < |\omega _{0}|^{2} < \omega _{c}^{2}$)
harmonic oscillator. 

The Gaussian nature of the system is a great simplification,
as demonstrated by eq.(\ref{def of influence functional}),
and one may write an effective action for the $q$ system
including the environment effect;
\begin{eqnarray}
i\,S_{{\rm eff}} &=&
i\,\int_{0}^{t}\,d\tau \,
(\,\frac{1}{2}\, \dot{\xi }\,\dot{X} + \frac{\omega _{0}^{2}}{2}\, \xi\,X\,)
\nonumber \\
&&
-\int_{0}^{t }d\tau \int_{0}^{\tau }ds
\left( \xi (\tau )\alpha_{R}(\tau - s)\xi (s) + i\,\xi (\tau )
\alpha _{I}(\tau - s)X(s) \right) \,.
\end{eqnarray}
The obvious linearity in the $X(\tau )$ variable here gives
a trivial $X(\tau )$ path integration in the form of a 
delta-functional; it determines the $\xi (\tau )$ path as
\begin{equation}
\frac{d^{2}\xi }{d\tau ^{2}} - \omega^{2} _{0}\,\xi (\tau ) +
2\,\int_{\tau }^{t}\,ds \,\xi (s)\,\alpha _{I}(s - \tau) = 0
\,.
\label{semiclassical eq fo xi} 
\end{equation}
The final $\xi-$path integration then leads to, as the exponent factor,
\begin{equation}
-\,
\int_{0}^{t}\,d\tau \,\int_{0}^{\tau }\,ds\,\xi (\tau )\alpha _{R}(\tau - s)
\xi (s) \,, 
\end{equation}
plus a surface term resulting after the $X(\tau )$ partial integration.
The function $\xi(\tau )$ here is the solution of 
eq.(\ref{semiclassical eq fo xi}) and 
the result of path integral
must be written with a specified boundary condition,
\( \:
\xi (0) = \xi _{i} \,, \hspace{0.3cm} \xi (t) = \xi _{f} \,.
\: \)

The standard technique of solving this type of integro-differential
equation for $\xi (\tau )$ is the Laplace transform
\cite{jmy-96-1}.
We shall only summarize the final result.
Two fundamental solutions to this equation are $g(t - \tau )$ and its
time derivative $\dot{g}(t - \tau )$ given by
\begin{eqnarray}
g(\tau ) 
&=&
\frac{N}{\omega _{B}}\,\sinh (\omega _{B}\tau )
+ 2\,\int_{\omega _{c}}^{\infty }\,d\omega \,H(\omega )\sin (\omega \tau )
\,,  \label{main g-formula} 
\\ &&
N = 1 - 
2\,\int_{\omega _{c}}^{\infty }\,d\omega \,\omega H(\omega )
< 1 \,.
\end{eqnarray}
The important properties are that $g(\tau)$ is odd and $\dot{g}(\tau)$
is even with $g(0) = 0$ and $\dot{g}(0) = 1$ which gives
the relation fixing $N$.
In terms of this basic function $g(\tau )$ a general solution
to the integro-differential equation with the given boundary 
condition is
\begin{eqnarray}
&&
\xi (\tau ) =
\frac{g(t - \tau )}{g(t)} \,\xi_{i} +
\left( \, \dot{g}(t - \tau ) - \frac{\dot{g}(t)}{g(t)}g(t - \tau )\,\right)
\,\xi _{f} \,.
\end{eqnarray}

The weight function $H$ here is  a discontinuity of some analytic
function ($F(z)$ of a complex variable $z = \omega $)
across the branch-point singularity
along the real axis at $\omega > \omega _{c}$,
and is given by
\begin{eqnarray}
&&
H(\omega ) = \frac{r(\omega )}{(\omega ^{2} + \omega _{0}^{2} -
\Pi(\omega ))^{2} + (\pi r(\omega ))^{2}}  \,,
\hspace{0.5cm} r(\omega ) = \frac{c^{2}(\omega )}{2\omega } \,, 
\\ &&
\Pi (\omega ) = {\cal P}\,\int_{\omega _{c}}^{\infty }\,d\omega' \,
\frac{2\omega '\,r(\omega' )}{\omega ^{2} - \omega'\, ^{2}} \,, 
\end{eqnarray}
where the integral for $\Pi $ stands for its principal part.
The value $\omega _{B}$ in eq.(\ref{main g-formula})
is determined as a solution for the isolated pole on the real axis at
\( \:
\omega^{2} = -\,\omega _{B}^{2} < 0 \,;
\: \)
\begin{eqnarray}
&&
-\,\omega _{B}^{2} + \omega _{0}^{2} + \int_{\omega _{c}}^{\infty }\,
d\omega \,\frac{2\omega\,r(\omega )}{\omega _{B}^{2} + \omega^{2}} = 0
\,.
\label{exact pole location} 
\end{eqnarray}
In general. one has
\begin{equation}
\omega _{B}^{2} > \omega _{0}^{2}
\,.
\end{equation}

Since many workers in this field use a renormalized potential
according to ref.\cite{caldeira-leggett 83}, we also introduce
these; 
\begin{eqnarray}
&&
V_{q} + V_{qQ} = V_{q}^{({\rm ren})}(q) + V_{qQ}^{({\rm ren})}(q \,, Q)  \,,
\\ &&
V_{q}^{({\rm ren})}(q) = V - 
q^{2}\,\int_{\omega _{c}}^{\infty }\,d\omega \,\frac{r(\omega )}{\omega } 
\equiv  -\,\frac{1}{2}\, \omega _{R}^{2}\,q^{2} \,, 
\\ &&
V_{qQ}^{({\rm ren})}(q \,, Q) = V_{qQ}(q\,, Q) +
q^{2}\,\int_{\omega _{c}}^{\infty }\,d\omega \,\frac{r(\omega )}{\omega }
\equiv V_{qQ}(q\,, Q) + \frac{1}{2}\, \delta \omega ^{2}\,q^{2} \,.
\end{eqnarray}
We shall redefine the $q-$system potential using $V_{q}^{({\rm ren})}(q)$.
This renormalized potential gives an inverted harmonic oscillator
with the curvature parameter $\omega _{R}$.
This renormalized curvature differs from the pole location of
the spectral function $H(\omega )$, namely $\omega _{B}$ by a factor
of order of the interaction coupling.
In the weak coupling limit these two quantities are given by
\begin{eqnarray}
&&
\omega _{R}^{2} 
=  \omega _{0}^{2} + 2\,\int_{\omega _{c}}^{\infty }\,d\omega \,
\frac{r(\omega )}{\omega } 
 \,,
\label{def of omega-r} 
\\ &&
\omega _{B}^{2} 
\approx \omega _{0}^{2} + 2\,\int_{\omega _{c}}^{\infty }\,d\omega \,
\frac{\omega\,r(\omega )}{\omega^{2} + \omega _{0}^{2}}  \,.
\end{eqnarray}
The equation for $\omega _{R}^{2}$ is a precise definition,
while the equation for $\omega _{B}^{2}$ is an approximate relation
valid for the weak coupling, the exact relation being given by
eq.(\ref{exact pole location}).
In general, one can prove that
\begin{equation}
\omega_B^2 < \omega_R^2 \,,
\end{equation}
beyond the weak coupling approximation.
As an example, in the Ohmic model which will be discussed shortly,
the relation becomes
\begin{equation}
\omega _{B} \approx \omega _{R} - \frac{1}{2}\,\eta \,,
\label{pole freq for ohmic model} 
\end{equation}
where $\eta $ is the Ohmic friction.
A more precise relation in this case is 
eq.(\ref{ohmic relation for b and r}).
The relation to the bare quantity $\omega _{0}$ is
\begin{equation}
\omega _{R} \approx \omega _{0} + k\, \eta \,,
\hspace{0.5cm}
k \approx \frac{\Omega}{\pi \omega_0} \,.
\end{equation}
In the infinite cutoff limit, $\Omega \rightarrow \infty\,$,
the quantity $k$ is divergent.

In Fig.1 we show schematically the analytic structure of
the function $F(z^{2})$, basic to the determination of the discontinuity
function $H(\omega )$.
Unlike the case of the normal harmonic oscillator for which 
the pole location may have an imaginary part, 
the pole at $z^{2} = -\,\omega _{B}^{2}$ appears exactly on the real axis.
The other singularity is the branch cut starting from
the threshold $z^{2} =\omega _{c}^{2}$.

The physical meaning of the basic function $g(\tau )$ is better
understood by solving the operator Langevin equation for this
system;
\begin{eqnarray}
&&
\frac{d^{2}q}{dt^{2}} - \omega _{0}^{2}\,q +
2\,\int_{0}^{t}\,d\tau \,\alpha _{I}(t - \tau )q(\tau ) =
F_{Q}(\tau ) \,. \label{langevin eq} 
\end{eqnarray}
The quantity $\omega_0^2$ here should be understood as a function
of $\omega_B^2$ eliminating $\omega_0^2$
with eq.(\ref{exact pole location}).
The homogeneous solution to this Langevin equation is given by
using $g(t)$ and $\dot{g}(t)$; with the initial
data of 
\( \:
q(0) \,, \; \dot{q}(0) = p(0) \,, 
\: \)
\begin{equation}
q(t) = q(0)\dot{g}(t) + p(0)g(t) \,.
\end{equation}
The main term 
\( \:
g(t) \approx N\,\sinh (\omega _{B}t)/\omega _{B}
\: \)
and $\dot{g}(t) \approx N\,\cosh (\omega _{B} t)$
describes an average motion $\langle q(t) \rangle$ under
the renormalized, inverted harmonic oscillator modified by environment,
for which the original parameter $|\omega _{0}^{2}|$
is replaced by the new shifted $\omega _{B}^{2}$.
Correction to this classical motion given by the second term 
in eq.(\ref{main g-formula}) describes a damped
oscillation with an amplitude decreasing by an inverse power of time
at large times .

A straightforward calculation of $X-$ and $\xi-$path integration
finally gives an effective action valid for 
any initial state of the $q-$system.
We first define new functions by
\begin{eqnarray}
&&
h(\omega \,, t) = \int_{0}^{t}\,d\tau \,g(\tau )\,e^{-i\omega \tau }
\,, \label{h-formula} 
\\ &&
k(\omega \,, t) = \int_{0}^{t}\,d\tau \,\dot{g}(\tau )\,e^{-i\omega \tau }
= i\omega h(\omega \,, t) + g(t)\,e^{-i\omega t}
\,.\label{k-formula} 
\end{eqnarray}
With the normalization fixed by unitarity,
one has for the effective action $J$ defined by
\( \:
\rho ^{(R)} = {\rm tr}\;(\,J \rho _{i}\,) \,, 
\: \)
\begin{eqnarray}
&&
J(X_{f} \,, \xi _{f} \; ; X_{i} \,, \xi _{i} \; ; t)
= 
\frac{1}{2\pi g(t)}\,e^{iS_{{\rm cl}}} \,, 
\end{eqnarray}
where $S_{{\rm cl}}$ is given by \cite{qbm review} 
\begin{eqnarray}
i\,S_{{\rm cl}} &=& -\,\frac{U}{2}\,\xi _{f}^{2} - \frac{V}{2}\,\xi _{i}^{2}
- W\,\xi _{i}\,\xi _{f}  + 
\frac{i}{2}\,X_{f}\,\dot{\xi }_{f} - 
\frac{i}{2}\,X_{i}\,\dot{\xi }_{i} \,, 
\\
U &=& 
(\frac{\dot{g}}{g})^{2}\,I_1 + I_2 - 2\frac{\dot{g}}{g}\,I_3
\,, \hspace{0.5cm}
V = \frac{1}{g^{2}}\,I_1 \,, 
\\ 
W &=& \frac{1}{g}\,I_3 - \frac{\dot{g}}{g^{2}}\,I_1
= \frac{1}{2}\,g\,\dot{V}
\,, 
\\
I_{1} &=& 
\int_{\omega _{c}}^{\infty }
\,d\omega \,\coth \frac{\beta \omega }{2}\,r(\omega )\,
|h(\omega \,, t)|^{2} \,, 
\label{i-1}
\\ 
I_{2} &=&
\int_{\omega _{c}}^{\infty }
\,d\omega \,\coth \frac{\beta \omega }{2}\,r(\omega )\,
|k(\omega \,, t)|^{2} \,, 
\label{i-2}
\\ 
I_{3} &=& 
\int_{\omega _{c}}^{\infty }
\,d\omega \,\coth \frac{\beta \omega }{2}\,r(\omega )\,
\Re [\,h(\omega \,, t)k^{*}(\omega \,, t)\,] \,, 
\label{i-3}
\\
\dot{\xi }(\tau ) &=&
-\,\xi _{i}\frac{\dot{g}(t - \tau )}{g(t)} - \xi _{f}\,
\left( \,\stackrel{..}{g}(t - \tau ) - \frac{\dot{g}(t - \tau )\dot{g}(t)}
{g(t)}\,\right) \,.
\end{eqnarray}

For the discussion of the barrier penetration factor,
we take a pure initial state given by a wave function
\( \:
\psi (x) \,.
\: \)
The Wigner function in this case is
\begin{eqnarray}
&&
f_{W}^{(R)}(x\,, p) = \int\,\frac{dx_{i}d\xi _{i}}{2\pi \sqrt{C}}\,
\psi ^{*}(x_{i} - \frac{1}{2}\, \xi _{i})
\psi (x_{i} + \frac{1}{2}\, \xi _{i})\,e^{-\,A}
\,,
\\ &&
A = \frac{{\rm det}\; I}{2C}\,
\left( \,\xi _{i} + i(g J_{1} + \dot{g}J_{3})(x - \dot{g}x _{i})
+ i(\dot{g}J_{2} + gJ_{3})(p - \stackrel{..}{g}x_{i})\,\right)^{2} 
\nonumber \\ &&
\hspace*{0.5cm} 
+\, \frac{1}{2}\, \left( \,J_{1}(x - \dot{g}x _{i})^{2} +
J_{2}(p - \stackrel{..}{g}x_{i})^{2}
+ 2J_{3}(x - \dot{g}x _{i})(p - \stackrel{..}{g}x_{i})
\,\right) \,, 
\\ &&
C = \int_{\omega _{c}}^{\infty }\,d\omega \,\coth \frac{\beta \omega }{2}
\,r(\omega )|\dot{g}h(\omega \,, t) - gk(\omega \,, t)|^{2} \,, 
\\ &&
(J) = (I)^{-1} \,, \hspace{0.5cm} 
J_{1\,, 2} = \frac{I_{2\,, 1}}{I_{1}I_{2} - I_{3}^{2}} \,, \hspace{0.5cm}
J_{3} = -\,\frac{I_{3}}{I_{1}I_{2} - I_{3}^{2}} \,. 
\end{eqnarray}

Although it is not used in the calculation of the flux factor in the
next subsection, it is of great theoretical interest 
to express our result as a transformation
of the initial Wigner function $f_{W}^{(i)}$ to the final one
$f_{W}^{(R)}$. We give this in Appendix A.
This mapping $f_{W}^{(i)} \rightarrow f_{W}^{(R)}$ is an integrated
form describing dynamics of the $q-$system.
Its differential form is known as the Fokker-Planck equation,
and we shall explain this briefly in Appendix B.

\vspace{0.5cm} 
\lromn 3B 
\hspace{0.2cm}
{\bf Barrier penetration factor}

\vspace{0.5cm} 
\hspace*{0.5cm} 
The flux at position $x$ is computed from the formula
\begin{equation}
I(x \,, t) = \int_{-\infty }^{\infty }\,\frac{dp}{\sqrt{2\pi }}\,p\,
f_{W}^{(R)}(x \,, p \; ; t)
\,,
\end{equation}
to give
\begin{eqnarray}
&& 
I(x \,, t) = \int\,\frac{dx_{i}d\xi _{i}}{2\pi g}\,
\psi ^{*}(x_{i} - \frac{\xi _{i}}{2})\psi (x_{i} + \frac{\xi _{i}}{2})\,
\left( \frac{\dot{g}}{g}x + (\stackrel{..}{g} - \frac{\dot{g}^{2}}{g})
x_{i} + iW\xi _{i} \right)\,
\nonumber \\ && \hspace*{0.5cm} 
\cdot \exp \left[ -\,\frac{V}{2}\left( \xi _{i} + \frac{i}{gV}\,
(x - \dot{g}x_{i})\right)^{2} - \frac{1}{2g^{2}V}\,
(x - \dot{g}x_{i})^{2}
\,\right]
 \,.
\end{eqnarray}

We use the WKB formula for energy eigenstates of IVHO.
Considering the incident left mover at $x < 0$ with
the unit flux, we take as the wave funtion 
at $x > x_{*}$ (the right turning point) 
\begin{eqnarray}
&&
\psi (x) \approx \frac{T(E)}{\sqrt{p(x)}}\,\exp 
\left( \,i\,\int_{x_{*}}^{x}\,dx'\,p(x')\,\right)
\,, \hspace{0.5cm} p(x) = \sqrt{\,2E + \omega _{B}^{2}\,x^{2}\,} \,,
\end{eqnarray}
where $x_{*} = \sqrt{2|E|}/\omega _{B}$ and
$T(E)$ is the transmission coefficient 
as a function of the energy $E$ in a pure quantum state.
This choice of the wave function gives the trasmission coefficient,
\begin{equation}
|T(E)|^{2} \approx \frac{1}{ 1 + e^{-\,2\pi E/\omega _{B}}} \,.
\end{equation}

A point which becomes important when we compare our result 
with those of other papers is how one prepares the initial state.
In many past works a thermal equilibrium between the subsystem and
the environment is often assumed, and in this context it is natural to
take for our choice of the pure state the reference system 
characterized by the curvature $\omega _{B}$, the exact pole curvature.
The choice of the WKB wave function, using
the curvature parameter $\omega _{B}$, fits with this picture.
But it should be kept in mind that
we may in principle take any reference curvature (hereafter denoted by 
$\tilde{\omega }$) and
in these cases we replace $\omega _{B}$ below by $\tilde{\omega }$.

Expanding wave functions in $\xi _{i}$, 
we derive a general formula for the dynamical factor.
This involves an infinite series of the expansion in $\xi$
of the initial density matrix element,
\begin{eqnarray}
  \psi(x_i + \xi_i/2)\psi^*(x_i - \xi_i/2)
  &=&
  \sum_{n = 0}^\infty
  {\cal A}_n(x_i)\xi_i^n
  \,.
\end{eqnarray}
The first few terms of this series are
\begin{eqnarray}
  &&
      {\cal A}_0(x_i)
       = |\psi(x_i)|^2, \hspace{0.5cm} 
      {\cal A}_1(x_i)
       = 
      \frac{1}{2}(\psi^*\partial\psi - \psi\partial\psi^*)(x_i),
      \\ &&
     {\cal A}_2(x_i)
      = 
      \frac{1}{8}
      \left[
        (\psi^*\partial^2\psi + \psi\partial^2\psi^*)(x_i)
        + |\partial\psi|^2(x_i)
      \right]
\,.
\end{eqnarray}

Computation of the flux factor is then given by
\begin{eqnarray}
&& \hspace*{1cm}
  I(x,t)
  =
  \sum_{n = 0}^{\infty} I_n(x,t) \,,
  \\ &&
  I_n(x,t)
  =
  \int\frac{dx_id\xi_i}{2\pi g}
  {\cal A}_n(x_i)\xi_i^n
  \left(
    \frac{\dot{g}}{g}x + (\ddot{g} - \frac{\dot{g}^2}{g})x_i
    + i\left(\frac{I_3}{g} - \frac{\dot{g}I_1}{g^2}\right)\xi_i
  \right)
  \nonumber \\ &&
  \times
  \exp
  \left[
    -\frac{V}{2}
    \left(
      \xi_i + \frac{i}{gV}(x - \dot{g}x_i)
    \right)^2
    - \frac{1}{2g^2V}(x - \dot{g}x_i)^2
  \right]
  \\ &&
  =
  \left(\frac{g}{\sqrt{I_1}}\right)^n
  \int \frac{d\alpha}{2\pi \dot{g}}\, {\cal A}_n
  \left(\frac{x}{\dot{g}} + \frac{\sqrt{I_1}}{\dot{g}}\alpha \right)
  e^{-\alpha^2/2}
  \nonumber \\ &&
  \hspace*{-1cm}
  \times
  \int d\beta\, (\beta + i\alpha)^n
  \left\{
    \frac{\ddot{g}}{\dot{g}}x +
    \left(
      \frac{\ddot{g}\sqrt{I_1}}{\dot{g}} - \frac{I_3}{\sqrt{I_1}}
    \right)\alpha
    + i\left(\frac{I_3}{g} - \frac{\dot{g}I_1}{g^2}\right)
    \frac{g}{\sqrt{I_1}}\beta
  \right\}e^{-\beta^2/2}
  \nonumber
  \,.
\end{eqnarray}
The second equality follows by a trivial scale change of
integration variables $x_{i} \,, \xi _{i}$.

One may use 
expansion of ${\cal A}_n(x/\dot{g} + \sqrt{I_1}\alpha/\dot{g}  )$
in powers of coupling,
\begin{eqnarray}
&& \hspace*{1cm} 
  {\cal A}_n
  \left(\frac{x}{\dot{g}} +  \frac{\sqrt{I_1}}{\dot{g}}\alpha  
  \right)
  =
  \nonumber 
\\ &&
\hspace*{-0.5cm} 
  {\cal A}_n
  \left(
    \frac{x}{\dot{g}}
  \right)
  +
  {\cal A}_n'
  \left(
    \frac{x}{\dot{g}}
  \right)
  \frac{\sqrt{I_1}}{\dot{g}}\alpha
  + \cdots + \frac{1}{(k - 1)\,!}\,
  {\cal A}_n^{(k - 1)}\left(\frac{x}{\dot{g}}\right)\,
    \left(
    \frac{\sqrt{I_1}}{\dot{g}}\alpha
  \right)^{k - 1}
  + \cdots
  \,.
\end{eqnarray}
For calculation of the penetration factor one only needs 
$x\longrightarrow \infty$ limit of the flux function.
From the WKB formula;
\begin{eqnarray}
  {\cal A}_n(x_i)
  =
  \frac{|T(E)|^2}{n!}
  \frac{d^n}{d\xi_i^n}
  \left.
    \frac{1}{\sqrt{p(x_i - \xi_i/2)p(x_i + \xi_i/2)}}
    \exp\left(i\int^{x_i + \xi_i/2}_{x_i - \xi_i/2}dx'\,p(x')\right)
  \right|_{\xi_i = 0}
  \,,
\end{eqnarray}
one has
\begin{eqnarray}
{\cal A}_n(x_i)
  \longrightarrow
  |T(E)|^2
  \left\{
    i\frac{(i\omega_Bx_i)^{n - 1}}{n!}
    + {\cal O}(x_i^{n - 3})
  \right\}
  \,.
\end{eqnarray}
Only the term containing $\left(\frac{\sqrt{I_1}}{\dot{g}}\alpha
  \right)^{n - 1}$ remains here.
One may then derive
\begin{eqnarray}
  I_n(x,t)
  \longrightarrow
    |T(E)|^2(-1)^n\left(\frac{\omega_Bg}{\dot{g}}\right)^{n - 1}
  \left(
    \frac{g\ddot{g}}{\dot{g}^2}  - 1
  \right)~,\qquad(n \geq 1)
  \,.
\end{eqnarray}
This along with 
\begin{equation}
I(\infty \,, t) =
|T(E)|^2\,f(t) \,,
\label{fundamental formula}
\end{equation}
gives a general formula for the dynamical factor,
\begin{eqnarray}
  f(t)
  =
  \frac{\ddot{g}}{\omega_B \dot{g}}
  -
   \left(
    \frac{g\ddot{g}}{\dot{g}^2}  - 1
  \right)
  \sum_{n = 1}^\infty
  \left(-\frac{\omega_Bg}{\dot{g}}\right)^{n - 1}
  =
  \frac{\ddot{g} + \omega_B\dot{g}}{\omega_B(\dot{g} + \omega_Bg)}
\,.
\label{general dynamical factor} 
\end{eqnarray}
This is the main result of the present work.

The salient feature of this result is factorization; 
the main suppression factor given by $|T(E)|^{2}$ is
affected by presence of
the thermal environment only via renormalization effect,
as will be more fully discussed shortly.
The other prefactor $f(t)$ carries dynamical information of 
the time evolution.

The basic functions $g(t)$ and $\dot{g}(t)$
that appear in the dynamical function
$f(t)$ are the solution to the homogeneous
part ($F_{Q} = 0$) of the Langevin equation (\ref{langevin eq})
with the initial condition,
\( \:
g(0) = 0 \,, \dot{g}(0) = 1 \,.
\: \)
More conveniently, one may rewrite the dynamical function as
\begin{equation}
f(t) = \frac{p_{0}(t)}{\omega _{B}\,q_{0}(t)} \,,
\hspace{0.5cm} 
p_{0}(t) = \dot{q}_{0}(t) \,.
\end{equation}
Here $q_{0}(t)$ is the solution with the initial condition,
\begin{equation}
\dot{q}(0) = \omega _{B}\,q(0) \,.
\end{equation}
This condition corresponds to the zero energy at time
$t=0$, since
\begin{eqnarray}
&&
H_{q} = \frac{1}{2}\, \dot{q}^{2} - \frac{1}{2}\, \omega _{B}^{2}\,q^{2}
= 0  \,.
\end{eqnarray}
Namely, the particle sits on the top of the renormalized
potential barrier.
Note however that the exact pole curvature $\omega _{B}$ appears
here instead of the renormalized $\omega _{R}$.
The ratio of the momentum to the coordinate
for $f(t)$ is a measure of the energy flow from the thermal environment.
Thus, the case of $f(t) > (<) 1$ corresponds to an energy 
inflow (outflow).
Both at $t\ll 1/\omega _{B}$ and $t \gg 1/\omega _{B}$
this function $f(t)$ is nearly unity and it deviates appreciably from
unity only within the time range of $1/\omega _{B}$.

An explicit formula useful for detailed analysis of the dynamical
function $f(t)$ is
\begin{eqnarray}
&&
\hspace*{-0.5cm}
q_{0}(t) = \dot{g}(t) + \omega _{B}\,g(t) =
N\,e^{\omega _{B}\,t} + 2\,
\int_{\omega _{c}}^{\infty }\,d\omega \,H(\omega )\,
\left( \, \omega \cos \omega t + \omega _{B} \sin \omega t \,\right)
 \,.
 \label{zero energy motion} 
\end{eqnarray}
The first term represents a simple classical motion under
the potential, modified by the curvature renormalization, while
the second term is a further deviation due to the environment interaction.
The environment effect for $f(t)$ appears in two ways;
the first via the definition of $\omega _{B}$ determined
by the potential renormalization due to the environment interaction.
The second one is the continuous part of spectral integral in
eq.(\ref{zero energy motion}), and its associated deviation of
the normalization factor $N$ from unity.

Before we go on to specific models of the environment, it
is appropriate to discuss some general results.
First, both in the weak coupling region and in the asymptotic
late time the dynamical function $f(t)$ behaves as
\begin{eqnarray}
&&
\hspace*{-1cm}
f(t) \approx f_{\rm{asym}}(t) \,, \hspace{0.5cm}
f_{\rm{asym}}(t) = 1 - \frac{2}{\omega _{B}N}\,e^{-\omega _{B}t}\,
\int_{\omega _{c}}^{\infty }\,d\omega \,H(\omega )\,
(\omega^{2} + \omega _{B}^{2})\,\sin \omega t \,.
\label{asymptotic f-formula} 
\end{eqnarray}
This form has the correct asymptotic behavior,
$f(\infty ) = 1$ as well as the correct initial behavior 
$f(0) = 1$ if the integral above is convergent.

On the other hand, one can prove for the initial-time behavior of $f(t)$;
\begin{eqnarray}
&&
f(t) = 1 - \omega _{B}\,(\,1 - 
\frac{\stackrel{...}{g}(0)}{\omega _{B}^{2}} \,)\,t + \cdots \,,
\label{initial-time behavior}
\\ &&
1 - \frac{\stackrel{...}{g}(0)}{\omega _{B}^{2}}  = 
\frac{2}{\omega _{B}^{2}}\,\int_{\omega _{c}}^{\infty }\,d\omega \,
H(\omega )\omega\,(\,\omega ^{2} + \omega _{B}^{2}\,) > 0 \,.
\label{coeff of small t}
\end{eqnarray}
The last inequality means that for very small $t \,, $ $f(t) < 1$,
giving rise to suppression for barrier penetration.
In order to have a convergent integral (\ref{coeff of small t}) 
for the formula of $\stackrel{...}{g}(0)$
a physical cutoff $\Omega $ of the environment spectrum is necessary, 
thus the simple-minded
Ohmic model without the cutoff should be treated with caution.

The asymptotic formula (\ref{asymptotic f-formula}), as will be shown
later, describes well, even  numerically,
the dynamical function at all times, including the small time.
For instance, the expansion in time $t$ of eq.(\ref{asymptotic f-formula})
gives almost the same result as eq.(\ref{initial-time behavior}),
except the missing factor $N$.

We would like to stress that two properties,
\( \:
f(t) < 1 
\: \)
with $f(0) = 1$ for small $t$ and
\( \:
f(\infty ) = 1 \,, 
\: \)
are generic features for any model having a finite physical
cutoff of the spectral weight $r(\omega )$.
As shown later, however, some approximations or models with the
infinite cutoff violate these general properties.

Although our fundamental formula (\ref{fundamental formula})
is derived for energy eigenstates,
one may compute the barrier penetration factor for any
mixed state by suitably weighting this rate for energy eigenstate.
A salient feature of our flux formula (\ref{fundamental formula})
is that the dependence on the initial (system) state is given
by the well known quantum mechanical barrier
penetration factor $|T(E)^{2}|$, the other factor $f(t)$ being
independent of the initial state.
This property of factorization is specific to the harmonic barrier.
For a more general potential barrier the dynamical factor
may depend on the energy of the initial state like
$f(t \,; E)$.

We would like to stress that our result extends the result of
\cite{caldeira-leggett 83} in several ways.
First, we derived the tunneling rate for any energy eigenstate, while
the authors of ref.\cite{caldeira-leggett 83} deals with the zero
temperature limit of the mixed state in complete equilibrium.
Our method is also completely different from the Caldeira-Leggett's
Euclidean approach, and our method
makes it possible to discuss dynamics of the time evolution.
The third point is that we derived the prefactor rigorously
unlike previous approximate calculations.
In the next section we explicitly show how we effectively
obtain the result of Caldeira and Leggett.

\vspace{0.5cm} 
\lromn 3C \hspace{0.2cm}
{\bf Some examples}
\vspace{0.5cm} 

We would like to compute this dynanical function for a few
typical examples of the environment spectrum. 
The spectral function $r(\omega )$ is taken as
\begin{eqnarray}
&&
(1)
\hspace{0.5cm} 
r_{O}(\omega ) = \frac{\eta}{\pi} \,\omega  \,,
\label{ohmic model} 
\\ &&
(2) \hspace{0.5cm} 
r_{D}(\omega ) = \frac{\eta \omega }{\pi\,(1 + \omega ^{2}/\Omega ^{2})} \,,
\\ &&
(3) \hspace{0.5cm} 
r_{T}(\omega ) = \frac{\eta}{\pi} \,
(\,\omega - \omega _{c}\,\epsilon (\omega )\,)
\,\theta (|\omega | - \omega _{c})\,
\theta (\Omega - |\omega |) \,,
\\ &&
(4) \hspace{0.5cm} 
r_{S}(\omega ) = \frac{\eta }{\pi }\,\omega ^{2}\,
\epsilon (\omega )\,\theta (\Omega - |\omega |) \,.
\end{eqnarray}
As usual, $\theta (x)$ is the step function and $\epsilon (x)$
the signature function.
The first one $r_{O}(\omega )$ is the Ohmic model (without physical
cutoff of the environment spectrum), the second
$r_{D}(\omega )$ the Drude model, and the fourth
$r_{S}(\omega )$ a super-Ohmic model, while the third one 
$r_{T}(\omega )$ has a threshold at $\omega _{c}$.
In Fig.2 these spectral weights are schematically depicted.
In the last three cases $\Omega $ acts as a cutoff of 
the environment spectrum. In Appendix C
we give parameters  necessary to compute the dynamical function.

We note here that the Ohmic model defined here by the spectral
weight $r_{O}(\omega )$ should, strictly speaking, be taken as some limit of
the infinite cutoff, This cutoff could be a straightforward
frequecncy cutoff like $|\omega | < \Omega $, or a smoother
one as in the case of the Drude model of a large $\Omega $.
The way how the cutoff is introduced does not matter 
provided a cutoff is there, but
in some evaluation of the integral one should first introduce
the cutoff and then take the infinite cutoff limit.

The dynamical factor $f(t)$ is plotted in Fig. 3$-$7 for these 
four cases.
As the friction $\eta $ becomes large, deviation of $f(t)$ from unity
becomes appreciable, but only in the time range of order
$1/\omega _{B}$.
For physical reasons we always take $\eta \leq \omega_B$.
We have found an interesting behavior of $f(t)$;
some models can give enhancement over the usual quantum formula 
in the time range $t \approx 1/\omega _{B}$.
These are the threshold model with a large $\omega _{c}$ and
the super-Ohmic model, for which the dynamical factor can exceed unity. 
The super-Ohmic model has also a peculiar feature that the dynamical
function $f(t)$ can even become negative for a short time interval.
The use of the asymptotic or the weak coupling expression for
$f_{\rm{asym}}(t)$, eq.(\ref{asymptotic f-formula}), is 
compared to the exact result in these figures.
Except at small initial times this approximate form gives
a reasonable fit to more exact results.
Presence of the normalization factor $N$ in the formula
is important to get a good agreement.

The initial time behavior of the Ohmic model (\ref{ohmic model}) does
not reflect the necessary condition of $f(0) = 1$,
since $f(t)$ at very small $t$ is singular having no smooth derivative
at $t = 0$ (the left and the right derivatives do not meet)
in the case of the infinite cutoff limit.
This implies that $f(0) < 1$ in this case indicates
anomaly associated with the infinite cutoff, and should not be
taken too seriously.
On the other hand, the Drude model for a large, but finite cutoff
has an expected decrease of $f(t)$ from unity at small times.
The larger the cutoff is, the more abrupt this decrease is, and
a local minimum of $f(t)$ appears at a time in proportion to $1/\Omega $.

Finally,
we note how the dynamical function behaves in the Ohmic, or the local
approximation.
We wish to distinguish this Ohmic approximation from the Ohmic model
we just discussed.
First, write eq.(\ref{langevin eq in ohmic approx}) for IVHO;
\begin{equation}
\frac{d^{2} q}{dt^{2}} + \eta \frac{dq}{dt}
- \omega _{R}^{2}\,q = F_{Q}(t) \,.
\end{equation}
There is a problem of how to interpret the zero energy solution
since $\omega _{B}$ we need for this is not well defined.
One choice might be to use the relation obtained from
$r_{O}(\omega )$ in the Ohmic model;
\begin{equation}
\omega _{B} = \sqrt{\omega _{R}^{2} + \frac{\eta ^{2}}{4}} 
- \frac{\eta }{2} \,.
\label{ohmic relation for b and r} 
\end{equation}
This relation is derived by using two exact definitions, the one
for $\omega _{B}$ (\ref{exact pole location}) and another for
$\omega _{R}$ (\ref{def of omega-r}), along with the form
of the weight $r_{0}(\omega )$.
In this case the dynamical function is trivial; $f(t) = 1$.
But it is by no means obvious that this choice is unique, since
without a specific form of the weight function there is no
way to locate the pole curvature $\omega_{B}$.

Another choice is to take a more phenomenological view limiting 
to the local Langevin equation, and
define the zero energy condition by taking $\omega _{R}$ 
for $\omega _{B}$; namely, at time 0,
\begin{equation}
H_{q} = \frac{1}{2}\, \dot{q}^{2} - \frac{1}{2}\, \omega_{R}^{2}\,
q^{2} = 0 \,.
\end{equation}
The zero energy solution that initially sits on the potential top is
then
\begin{equation}
q_{0}(t) = N\,\left( \,(\omega_{+} + \omega _{R})\,e^{\omega _{-}t}
+ (\omega _{-} - \omega _{R})\,e^{-\omega _{+}t} \,\right) \,.
\end{equation}
The parameters here are given by
\begin{eqnarray}
&&
\omega _{\pm } = \sqrt{\omega _{R}^{2} + \frac{\eta ^{2}}{4}} 
\pm \frac{\eta }{2}
 \,,
\end{eqnarray}
both of which are larger than $\omega _{R}$.
It is then easy to get
\begin{eqnarray}
&&
f(t) = \frac{\omega _{-}(\omega _{+} + \omega _{R})\,e^{\omega _{-}t}
- \omega _{+}(\omega _{-} - \omega _{R})\,e^{-\omega _{+}t}}
{\omega_{R}\,\left( \,
(\omega _{+} + \omega _{R})\,e^{\omega _{-}t} +
(\omega _{-} - \omega _{R})\,e^{-\omega _{+}t}
\,\right)}
\,.
\label{ohmic aprox for f} 
\end{eqnarray}
This function behaves reasonably at initial times, having
$f(0) = 1$. More precisely,
\begin{equation}
f(t) \rightarrow 1 - \eta\,t \,.
\end{equation}
Its asymptotic late time behavior is given by
\begin{eqnarray}
&&
f(t ) \rightarrow f(\infty)\,(\,
1 + C\,e^{-\,(\omega_{+} + \omega _{-})\,t } \,)  
\,,
\\ &&
f(\infty) = \frac{\omega _{-}}{\omega _{R}} < 1
\,, \hspace{0.5cm}
C = \frac{(\omega_+ + \omega_-)(\omega_R - \omega_-)}
{\omega_-\,(\omega_+ + \omega_R)} > 0 \,.
\end{eqnarray}
The function $f(t)$ of (\ref{ohmic aprox for f}) is plotted in
Fig.8.
The property $f(\infty ) < 1$ shows an anomalous behavior
of this local approximation, which we regard as an defect
inherent in the local approximation.

\vspace{1cm}

\lromn 4 \hspace{0.2cm}
{\bf Applications}

\vspace{0.5cm} 
To illustrate advantages of our approach, we take up two applications of
our basic formula, eq.(\ref{fundamental formula}) along with
(\ref{general dynamical factor}).
The first example is computation of the tunneling rate for
the kind of potential depicted in Fig. 9.
Qualitatively, a new normal harmonic oscillator is added in the
left region of the previous inverted harmonic oscillator.
The simplest example of this class of potential is a cubic form,
\begin{equation}
V(x) = -\,\frac{\omega _{B}^{3}}{3\sqrt{6V_{0}}}\,x^{3}
- \frac{\omega _{B}^{2}}{2}\,x^{2} \,.
\label{cubic potential}
\end{equation}
Here $V_{0}$ is the barrier height measured from the left bottom
of the potential, and $\omega _{B}$ is taken real and positive.
The important quantity added, a new frequency
$\omega _{*}$ in the left harmonic well, is equal to $\omega _{B}$
in this cubic potential.
Below we generally distinguish the two, $\omega _{B}$ and $\omega _{*}$,
having a more general tunneling potential in mind.

The problem we set up is to compute
the tunneling rate trapped in the metastable
state in the left oscillator part at temperature $T$, the same
temperature as the environment.
The setting of this problem is the same as in the Euclidean
approach of Grabert et al.\cite{qbm review}.
We assume that the temperature $T \leq V_{0}$ 
with a further condition of $V_{0} \gg \omega _{*}$.
Under this circumstance the $n$-th energy eigenstate of the left
oscillator is distributed with the probability $w_{n}$ of
the Boltzmann-Gibbs ensemble,
\begin{equation}
w_{n} = \frac{e^{-\,n\,\beta \omega _{*}}}
{\sum_{m}\,e^{-\,m\,\beta \omega _{*}}} =
\left( 1 - e^{-\,\beta \omega _{*}} \right)\,e^{-\,n\,\beta \omega _{*}}
\,.
\end{equation}
We then convolute with this weight the barrier penetration factor,
to get the tunneling probability $\Gamma $,
\begin{eqnarray}
&&
\Gamma(T \,, t) = \left( 1 - e^{-\,\beta \omega _{*}} \right)\,
\sum_{n=0}^{\infty }\,e^{-\,n\,\beta \omega _{*}}
\,|T(\,-\,V_{0} + \omega _{*}(n + \frac{1}{2}\, )\,)|^{2}
\,f(t)
 \,.
\label{finite T penetration}
\end{eqnarray}

We first discuss the infinite time limit, in which
\( \:
f(t) \rightarrow 1 \,.
\: \)
This probability has the zero temperature limit,
\begin{equation}
\Gamma (0 \,, \infty ) \approx |T(-\,V_{0} + \frac{\omega _{*}}{2})|^{2} =
\frac{1}{1 + e^{2\pi (V_{0} - \frac{\omega _{*}}{2})/
\omega_B}} \,, 
\end{equation}
which is valid for $T \ll \omega _{*}$.
On the other hand, for $T \gg \omega _{*}$ the discrete sum
\begin{equation}
\Gamma (T \,, \infty ) = 
2\,\sinh \frac{\beta \omega_*}{2}\,e^{-\beta V_{0}}\,
\sum_{n= 0}^{\infty }\,\frac{e^{-\,\beta \omega_*\,
(n + \frac{1}{2} - V_{0}/\omega_*)}}
{1 + e^{-\,2\pi \,(\,(n + \frac{1}{2})\omega_* - V_{0}\,)/\omega_B}}
\,,
\end{equation}
holds. This approximately reduces to
\begin{eqnarray}
&&
\Gamma (T \,, \infty ) 
  \approx 
  2\frac{\beta\omega_*}{2}\,e^{-\beta V_0}
  \int^\infty_0
  \frac{dx}{\beta\omega_*}
  \frac{1}{1 + x^{2\pi/\beta\omega_B}}
  \nonumber \\   &&
\\ && \hspace*{1cm} 
  = \frac{\beta\omega_B\,e^{-\beta V_0}}
  {2\sin(\beta\omega_B/2)}
  \approx e^{-\beta V_0} \,.
\end{eqnarray}
The last formula holds for $T \gg \omega_B$.
This is the expected classical formula for the barrier penetration
at finite temperature.
More quantitatively we numerically computed eq.(\ref{finite T penetration}) 
to compare with various approximate formulas.
The quantity $\Gamma (T \,, \infty )$, 
computed from eq.(\ref{finite T penetration}) with $f(t) = 1$, is
plotted in Fig.10.
In this figure an approximate formula,
\begin{equation}
  \Gamma(T,\infty) =
  \frac{\omega_B\sinh(\beta\omega_*/2)}{\omega_*\sin(\beta\omega_B/2)}
    \,e^{-\beta V_0}
  + \frac{1}{1 + e^{-\,2\pi(\omega_*/(2\omega_B) - V_0/\omega_B)}} \,, 
\label{approximate probability for finite t}
\end{equation}
is compared to the exact result in the case of $\omega_B = \omega_*$.
This interpolation formula is a simple sum of the improved high and
the improved low temperature limit.

We now present our understanding of the result of Caldeira and Leggett
for the cubic potential $V(x)$ (\ref{cubic potential}).
For this we extend our result of IVHO to this class of potential by
a new formula,
\begin{equation}
I(\infty \,, t) =
\exp\left(\,-\,2\,\int_{x_1}^{x_2}\,dx\,\sqrt{2\,(V(x) - E)}\,
\right) \,
\frac{1}{\omega _{B}}\,\frac{d}{dt}\,\ln |q_{0}(t)| \,,
\end{equation}
where $x_i$ are turning points for the energy $E = -\,V_0$ 
solution.
We take as the reference curvature $\tilde{\omega } = \omega _{B}$.
The trajectory function $q_{0}(t)$ is taken as the classical, zero energy
solution, but we ignore the dynamical function since it
is almost unity in the equilibrium circumstance of \cite{caldeira-leggett 83}.
Thus we find for the tunneling probability to be compared
\begin{equation}
I(\infty \,, t)  \approx |T(E)|^{2} = 
\exp\left(\,-\,\frac{36}{5}\,\frac{V_0}{\omega_B}\,\right) \,.
\end{equation}

The authors of \cite{caldeira-leggett 83} write the tunneling
probability in terms of the renormalized curvature $\omega_R$ and 
the friction $\eta$. 
For the Ohmic model of small friction, we get
(using eq.(\ref{ohmic relation for b and r}) with $\eta \ll
\omega_R$),
\begin{equation}
|T(E)|^{2} \approx e^{-\,2\pi V_0/\omega_R}\,e^{-\,\Delta B}
\end{equation}
with
\begin{equation}
\Delta B \approx  \frac{18}{5}\,\frac{\eta V_{0}}{\omega _{R}^{2}} 
=3.6\,\frac{\eta V_{0}}{\omega _{R}^{2}} 
\,.
\end{equation}
If we had used the IVHO potential instead of the cubic form,
we would have obtained a numerical factor slightly different
from this value, $3.6 \longrightarrow \pi \approx 3.14$.
On the other hand, Caldeira and Leggett obtain, using a different method,
\begin{equation}
\Delta B_{{\rm CL}} \approx \frac{162\,\zeta (3)}{\pi ^{3}}\,
\frac{\eta V_{0}}{\omega _{R}^{2}} 
\approx 6.28045 \,\frac{\eta V_{0}}{\omega _{R}^{2}} 
\,.
\end{equation}
Thus our result gives a result numerically different by a factor of
about 2 in the weak coupling limit.

On the other hand, in the large coupling limit, $\eta \gg 2\omega _{R}$,
these two are
\begin{equation}
\Delta B = \frac{36}{5} \,\frac{\eta V_{0}}{\omega _{R}^{2}} \,,
\hspace{0.5cm} 
\Delta B_{{\rm CL}} = 3\pi \,\frac{\eta V_{0}}{\omega _{R}^{2}}
\,.
\end{equation}
(Although our notation here suggests that this is a correction term
to the main term of $e^{-\,2\pi V_{0}/\omega _{R}}$, it is actually
a leading term for the case of $\eta \gg 2\omega _{R}$.)
Our result is about 0.8 times the Caldeira-Leggett value.

In general, when one writes the deviation of the penetration
factor from the case of no environment 
as $e^{-\,\Delta B}$, one has the form,
\begin{equation}
\Delta B  = \Phi\left(\,\frac{\eta}{2\omega_R}\,\right)\,
\frac{\eta\, V_0}{\omega_R^2}
\,.
\end{equation}
Both in our case and in Caldeira and Leggett
the function $\Phi(\alpha)$ is slowly varying albeit numerically 
different, and
the deviation factor is dominated by $\eta\, V_0/\omega_R^2$.
This appears to be the most important dependence of parameters,
the numerical factor being a secondary effect.

In our interpretation of the result of Caldeira and Leggett
it is crucial to use the pole curvature
$\omega _{B}$ as our reference, and this choice 
is reasonable, because it corresponds to an equilibrium 
in the zero temperature limit considered in \cite{caldeira-leggett 83}.
With the dynamical function taken as $f(t) =1$,
this is the only way how the friction ($\eta$) dependence
can appear in our approach, namely via the parameter $\omega_B$
in the initial density matrix.
It is important that our inequality $\omega_B^2 < \omega_R^2$
implies the general result of suppressed rate of tunneling in
medium, the main point stressed by Caldeira and Leggett.
We also note that detailed comparison between our result and that of
\cite{caldeira-leggett 83} is possible only by assuming the relation
(\ref{ohmic relation for b and r}), specific to the Ohmic
spectrum.
For different models the difference between the two
might be larger.

In future we wish to examine further this discrepancy after we
develop formalism for more general potentials.

The second application is the problem of time evolution for the
same potential as above.
For this we consider the temperature range of
$\omega _{*} \ll T \ll V_{0}$ such that the average energy
eigenstate (of $\overline{E} \approx T$)
may be treated semiclassically.
We thus regard a particle in the $n$-th energy level ($n \gg 1$)
of the left harmonic oscillator as moving almost classically
with the periodic motion of frequency $\omega _{*}$.
Each time this particle hits against the right barrier,
it has a prescribed probability (\ref{fundamental formula})
of tunneling into the far right region.
Starting at time $t = 0$, one counts the $k$-th encounter with the barrier
at time $t = 2\pi k/\omega _{*}$ until the final time $t_{f}$ such that
there are roughly $t_{f}\omega _{*}/2\pi $ times of
possibilities of the barrier penetration.
We find it reasonable to use a reset time for each encounter
for the time $t$ of eq.(\ref{fundamental formula}).
Summing up all these possibilities, 
one gets the total tunneling probability,
\begin{eqnarray}
&&
\Gamma_{n} (t_{f}) =
|T(E_{n})|^{2}\,\frac{t_{f}\,\omega _{*}}{2\pi }
\,\overline{f} \,, 
\end{eqnarray}
where $\overline{f}$ is some sort of representative value for the
dyanamical factor for each encounter, perhaps some average of $f(t)$ 
(\ref{general dynamical factor}) over one period of
oscillation under the harmonic well like
\begin{equation}
\overline{f} = \frac{1}{t_{*}}\,\int_{0}^{t_{*}}\,dt\,f(t) \,, \hspace{0.5cm} 
t_{*} = \frac{\omega _{*}}{2\pi } \,.
\label{average f} 
\end{equation}
Another choice for $\overline{f}$ is the dynamical function at some particular
value of time during one period of oscillation, 
for instance at the classical turning point.

For $\omega _{B} \gg \omega _{*}\,, $  
$\overline{f} \approx 1$, and 
\begin{equation}
\Gamma_{n} (t) \approx |T(E_{n})|^{2}\,\frac{t\,\omega _{*}}{2\pi }\,.
\end{equation}
Thus, the total probability $\propto $ time $t$, and one may
define the tunneling rate per unit time,
\begin{equation}
\frac{\Gamma_{n} (t)}{t} =
\frac{\omega_{*}}{2\pi}\, |T(E)|^2 \,.
\end{equation}
This is nothing but the classic Kramers formula \cite{kramers}.

On the other hand, for $\omega _{B} \approx  \omega _{*}$
there may be a large deviation from the quantum mechanical formula.
We plot in Fig.11 some examples of the factor $\overline{f}$ 
(\ref{average f}) as a function of the average time $t_{*}$.
In most cases studied, $\overline{f}$ is some fraction of unity,
typically larger than 0.5.

\newpage
{\bf Appendix A 
\hspace{0.2cm} Integral transform of the Wigner function}

\vspace{0.5cm} 
After some algebra, we obtain
\begin{eqnarray}
&& 
f_{W}^{(R)}(x\,, p) = \frac{1}{2\pi \,\sqrt{I_{1}I_{2} - I_{3}^{2}}}
\,\int_{-\infty }^{\infty }\,\,dx_{i}\,
\int_{-\infty }^{\infty }\,dp_{i}\,f_{W}^{(i)}(x_{i}\,, p_{i})
\nonumber \\ && \hspace*{-1cm}
\cdot \exp \left[ -\,\frac{1}{2(I_{1}I_{2} - I_{3}^{2})}\,
\left( I_{1}(p - p_{{\rm cl}})^{2} + I_{2}(x - x_{{\rm cl}})^{2}
- 2I_{3}(x - x_{{\rm cl}})(p - p_{{\rm cl}})\right)\right]
\,, \label{wigner transform} 
\\ && \hspace*{1cm} 
x_{{\rm cl}} = \dot{g}x_{i} + gp_{i} \,, \hspace{0.5cm} 
p_{{\rm cl}} = \,\stackrel{..}{g}x_{i} + \dot{g}p_{i} \,.
\end{eqnarray}
The definitions of $I_i$ are given in the text,
eqs.(\ref{i-1})$-$(\ref{i-3}).
The time dependent functions, $x_{{\rm cl}}(t) \,, \; p_{{\rm cl}}(t)$,
are homogeneous solutions to the Langevin equation (\ref{langevin eq})
with $F_{Q} = 0$.

One may view the mapping from $f_{W}^{(i)}$ to $f_{W}^{(R)}$ as 
a kind of fluid flow. 
Compared to the classical mapping given in Section \lromn 2, 
the quantum solution (\ref{wigner transform})
in thermal medium is not deterministic with a broadening
given by the coefficient matrix,
\begin{equation}
(I) = 
\left( \begin{array}{cc}
I_{1} & I_{3} \\
I_{3} & I_{2}
\end{array}
\right)
\,.
\end{equation}
Moreover, the initial distribution $f_{W}^{(i)}$ itself is broadened
by quantum mechanical effects.
The peak point of the distribution is at
\( \:
\left( x_{{\rm cl}}(t) \,, p_{{\rm cl}}(t) \right) \,.
\: \)
One might imagine that the mapping 
\( \:
(x_{i} \,, p_{i}) \rightarrow 
(x_{{\rm cl}}(t) \,, p_{{\rm cl}}(t))
\: \)
is not invertible due to dissipative effects from the environment. 
This is not true; the mapping is actually invertible and
\begin{eqnarray}
&&
x_{i} = 
\frac{\dot{g}x_{{\rm cl}} - gp_{{\rm cl}}}
{\dot{g}^{2} - g\stackrel{..}{g}} \,, 
\hspace{0.5cm} 
p_{i} =
\frac{-\stackrel{..}{g}x_{{\rm cl}} + \dot{g}p_{{\rm cl}}}
{\dot{g}^{2} - g\stackrel{..}{g}} 
\,, 
\end{eqnarray}
with $\dot{g}^{2} - g\stackrel{..}{g} \neq 0$.

\vspace{1cm}

{\bf Appendix B 
\hspace{0.2cm} Fokker-Planck equation}

\vspace{0.5cm} 
One may derive the master equation for the reduced density
matrix as described in ref.\cite{master eq for ho}.
Our formula for the Wigner function (\ref{wigner transform}) is considered as
an explicit and convenient solution to this type of master equation.
From the master equation one can derive a Fokker-Planck equation
for the Wigner function;
\begin{eqnarray}
\frac{\partial f_{W}^{(R)}}{\partial t} 
&=&
- \,p\frac{\partial f_{W}^{(R)}}{\partial x} + \Omega ^{2}(t)\,
x\frac{\partial f_{W}^{(R)}}{\partial p} + C(t)\,\frac{\partial }{\partial p}
(pf_{W}^{(R)}) - 2D_{pp}(t)\,\frac{\partial ^{2}f_{W}^{(R)}}{\partial p^{2}}
\nonumber \\
&+& 2D_{xp}(t)\,\frac{\partial ^{2}f_{W}^{(R)}}{\partial x\partial p}
\,, 
\\
\Omega ^{2}(t) &=&
-\,\frac{\stackrel{..}{g}^{2} - \dot{g}\stackrel{...}{g}}
{g\stackrel{..}{g} - \dot{g}^{2}} =
\frac{\dot{g}}{g}\,\frac{d}{dt}\ln \left( \,\frac{g}{\dot{g}}\,
(\,\stackrel{..}{g} - \frac{\dot{g}^{2}}{g}\,)\,\right) \,, 
\\
C(t) &=& -\,\frac{g\stackrel{...}{g} - \dot{g}\stackrel{..}{g}}
{g\stackrel{..}{g} - \dot{g}^{2}} \,, 
\\
D_{pp}(t) &=&
\frac{1}{2}\, \left( \,\frac{g\stackrel{...}{g} - \dot{g}\stackrel{..}{g}}
{g\stackrel{..}{g} - \dot{g}^{2}}\,U
+ \frac{\dot{g}}{2g}\,\frac{g^{2}\stackrel{...}{g} - 
2g\dot{g}\stackrel{..}{g} + \dot{g}^{3}}{g\stackrel{..}{g} - \dot{g}^{2}}
\,W - \frac{\dot{U}}{2} - \dot{g}\dot{W}\,\right) \,, 
\\
D_{xp}(t) &=&
U - g\dot{W} + \frac{g^{2}\stackrel{...}{g} - 
2g\dot{g}\stackrel{..}{g} + \dot{g}^{3}}{g\stackrel{..}{g} - \dot{g}^{2}}\,W
\,,
\end{eqnarray}
where coefficient functions, 
\( \:
\Omega ^{2}(t) \,, \;C(t) \,, \; D_{pp}(t) \,, \; D_{xp}(t)
\: \)
are local functions of time and are written in terms of
$g(t)\,, \; U \,, \; V\,, \; W$.

Quantities that appear in this equation are well understood by
writing a set of moment equations of low orders;
\begin{eqnarray}
\frac{d\langle x \rangle}{dt} &=&
\langle p \rangle \,, \hspace{0.5cm} 
\frac{d\langle p \rangle}{dt} = - \,\Omega ^{2}(t)\,\langle x \rangle
- C(t)\langle p \rangle \,, 
\\
\frac{d\langle x^{2} \rangle}{dt} &=&
2\,\langle xp \rangle \,, 
\\
\frac{d\langle p^{2} \rangle}{dt} &=&
-\,2\Omega ^{2}(t)\,\langle xp \rangle
- 2C(t)\langle p^{2} \rangle - 4D_{pp}(t) \,, 
\\
\frac{d\langle xp \rangle}{dt} &=&
\langle p^{2} \rangle - \Omega ^{2}(t)\langle x^{2} \rangle
- C(t)\langle xp \rangle + 2D_{xp}(t) \,.
\end{eqnarray}
For instance, the quantity $\Omega ^{2}(t)$ here is a time dependent
curvature parameter  modified from the original 
$\omega _{0}^{2} $ to that in thermal medium, while
$C(t)$ is a time dependent friction.
In similar fashons one understands $D_{pp}$ and $D_{xp}$ as
fluctuations.
Physical behaviors of the harmonic oscillator system under
thermal environment are all determined by these four quantities
which are functions of the local time $t$.

Limiting values relevant to large times, $t \gg 1/\omega _{B}$, are
\begin{eqnarray}
&&
\Omega ^{2}(t) \approx -\,\omega _{B}^{2} \,, \hspace{0.5cm} 
C(t) \approx 0
 \,,
\\ &&
D_{pp}(t) \approx -\,\frac{\omega _{B}}{4}\,\dot{g}W \approx 
\frac{\omega _{B}^{3}}{4}\,
\int_{\omega _{c}}^{\infty }\,d\omega \,
\cosh (\frac{\beta \omega }{2})\,H(\omega ) \,, 
\\ &&
D_{xp}(t) \approx U - \dot{g}W \approx 
\int_{\omega _{c}}^{\infty }\,d\omega \,
\cosh (\frac{\beta \omega }{2})\,(\omega ^{2} + \frac{5}{4}\omega _{B}^{2})
\,H(\omega ) 
\,.
\end{eqnarray}

\vspace{1cm}

{\bf Appendix C 
\hspace{0.2cm} Parameters in specific environment models}

\vspace{0.5cm} 
\hspace*{0.5cm} 
We give various parameters in four specific models of environment given
in Section \lromn 3C. 
These parameters are used to calculate $g(t)$ according to
\begin{eqnarray}
  g(t)
  &=&
  \frac{N}{\omega_B}\sinh(\omega_B t)
  +
  2\int^\infty_{\omega_c} d\omega H(\omega)\sin(\omega t) \,,
   \\
  H(\omega)
  &=&
  \frac{r(\omega)}
  {(\omega^2 + \omega_R^2 - \Pi(\omega))^2 + \pi^2r(\omega)^2} \,,
  \\
  \omega_R^2
  &=&
  \omega_B^2 + C(\omega_B) \,.
\end{eqnarray}

The parameters are given as follows.

(1) Ohmic model
\begin{eqnarray}
  N
  =\left(1 + \frac{\eta}{2\omega_B}\right)^{-1},
  \qquad
  C(\omega_B)
  =
  \eta\omega_B,
  \qquad
  \Pi(\omega)
  =
  0 \,,
\end{eqnarray}
\begin{eqnarray}
  g(t)
  =
  \frac{N}{2\omega_B}
  \left(
    e^{\omega_B t}
    -
    e^{-(\omega + \eta)t}
  \right) \,.
\end{eqnarray}

(2) Drude model
\begin{eqnarray}
  N
  &=&
  \left(
    1 + \frac{\eta}{2\omega_B}
    \frac{\Omega^2}{(\omega_B + \Omega)^2}
  \right)^{-1},
  \\
  C(\omega_B)
  &=&
  \frac{\eta\omega_B\Omega}{\omega_B + \Omega},
  \qquad
  \Pi(\omega)
  =
  \frac{\eta\omega^2\Omega}{\omega^2 + \Omega^2} \,,
\\
  g(t)
  &=&
  \frac{N}{\omega_B}\sinh(\omega_B t)
  +
  \frac{\eta\Omega^2}
  {(\alpha_+^2 - \omega_B^2)
    (\alpha_-^2 - \omega_B^2)(\alpha_+^2 - \alpha_-^2)}
   \nonumber \\
  &\times&
  \left(
    (\alpha_+^2 - \alpha_-^2)e^{-\omega_Bt}
    +
   (\alpha_-^2 - \omega_B^2)e^{-\alpha_+t}
   +
   (\omega_B^2 - \alpha_+^2)e^{-\alpha_-t}
  \right) \,,
  \\
  \hspace*{1cm}
  \alpha_\pm
    &=&
  \frac{\omega_B + \Omega}{2}
  \pm
  \sqrt{
    \frac{(\omega_B - \Omega)^2}{4}
    -
    \frac{\eta\Omega^2}{\omega_B + \Omega}
    } \,.
\end{eqnarray}

(3) Threshold model
\begin{eqnarray}
  N
  &=&
  \left(
    1
    -\frac{\eta}{\pi}\frac{\Omega - \omega_c}{\Omega^2 + \omega_B^2}
    + \frac{\eta}{\pi\omega_B}\arctan\frac{\Omega}{\omega_B}
    - \frac{\eta}{\pi\omega_B}\arctan\frac{\omega_c}{\omega_B}
  \right)^{-1} \,,
  \\
  C(\omega_B)
  &=&
  \frac{2\eta\omega_B}{\pi}
  \left(
    \arctan\frac{\Omega}{\omega_B}
    -
    \arctan\frac{\omega_c}{\omega_B}
  \right)
  \nonumber \\
  &+&
  \frac{\eta\omega_c}{\pi}
  \left\{
    \ln\left(1 + \frac{\omega_B^2}{\Omega^2}\right)
    -
    \ln\left(1 + \frac{\omega_B^2}{\omega_c^2}\right)
  \right\} \,,
  \\
  \Pi(\omega)
  &=&
  \frac{\eta(\omega - \omega_c)}{2\pi}
  \ln\left(\frac{\omega}{\omega_c} - 1\right)^2
  -
  \frac{\eta(\omega + \omega_c)}{2\pi}
  \ln\left(\frac{\omega}{\omega_c} + 1\right)^2
  \nonumber \\
  &-&
  \frac{\eta(\omega - \omega_c)}{2\pi}
  \ln\left(\frac{\omega}{\Omega} - 1\right)^2
  +
  \frac{\eta(\omega + \omega_c)}{2\pi}
  \ln\left(\frac{\omega}{\Omega} + 1\right)^2 \,.
\end{eqnarray}

(4) Super-Ohmic model
\begin{eqnarray}
  N
  &=&
  \left(
    1 - \frac{\eta}{\pi}\frac{\Omega^2}{\Omega^2 + \omega_B^2}
    + \frac{\eta}{\pi}\ln\frac{\Omega^2 + \omega_B^2}{\omega_B^2}
  \right)^{-1} \,,
   \\
  C(\omega_B)
  &=&
  \frac{\eta\omega_B^2}{\pi}\ln\frac{\Omega^2 + \omega_B^2}{\omega_B^2},
  \qquad
  \Pi(\omega)
  = - \frac{\eta \omega^2}{\pi}\ln\frac{\Omega^2 - \omega^2}{\omega^2}
  \,.
\end{eqnarray}

In calculation of $g(t)$ for the Ohmic model one needs a frequency 
cutoff in intermediate steps of integration, but the final result 
does not depend on this cutoff factor. For the model having a 
threshold and for the suer-Ohmic model we cannot get analytic forms 
of the basic function $g(t)$.

\newpage

\vspace{1cm}
\begin{center}
{\bf Acknowledgment}
\end{center}

The work of Sh. Matsumoto is partially
supported by the Japan Society of the Promotion of Science.

\vspace{1cm} 

\newpage

\begin{Large}
\begin{center}
{\bf Figure caption}
\end{center}
\end{Large}

\vspace{0.5cm} 
\hspace*{-0.5cm}
{\bf Fig.1}

Analytic structure giving the spectrum.
The pole at $-\,\omega_B^2$ moves as indicated when
the friction becomes large with $\omega_R^2$ fixed.

\vspace{0.5cm} 
\hspace*{-0.5cm}
{\bf Fig.2}

Schematic form of model spectral weights.

\vspace{0.5cm} 
\hspace*{-0.5cm}
{\bf Fig.3}

Dynamical function for the Ohmic model.
Values of the friction relative to the system curvature
are given for each case.
Dotted lines are calculated using the approximate, asymptotic formula
$f_{\rm{asym}}(t)$, 
eq.(\ref{asymptotic f-formula}), in the text.

\vspace{0.5cm} 
\hspace*{-0.5cm}
{\bf Fig.4}

Dynamical function for the Drude model.
Values of the friction and the cutoff relative to the system curvature
are given for each case.
Dotted lines are calculated using  the approximate, asymptotic formula
$f_{\rm{asym}}(t)$,
eq.(\ref{asymptotic f-formula}),
in the text.

\vspace{0.5cm} 
\hspace*{-0.5cm}
{\bf Fig.5}

Dynamical function for the threshold model of low
threshold.
Values of the friction, the cutoff, and the threshold
relative to the system curvature are given for each case.
Dotted lines are calculated using  the approximate, asymptotic formula
$f_{\rm{asym}}(t)$,
eq.(\ref{asymptotic f-formula}),
in the text.

\vspace{0.5cm} 
\hspace*{-0.5cm}
{\bf Fig.6}

The same as in {\bf Fig.5},
for the threshold model of high threshold.

\vspace{0.5cm} 
\hspace*{-0.5cm}
{\bf Fig.7}

Dynamical function for the super-Ohmic model.
Values of the friction and the cutoff are given for each case.
Dotted lines are calculated using  the approximate, asymptotic formula
$f_{\rm{asym}}(t)$,
eq.(\ref{asymptotic f-formula}),
in the text.

\vspace{0.5cm} 
\hspace*{-0.5cm}
{\bf Fig.8}

Dynamical function in the local, Ohmic approximation.
Values of the friction are given for each case.

\vspace{0.5cm} 
\hspace*{-0.5cm}
{\bf Fig.9}

Schematic form of a general potential.

\vspace{0.5cm} 
\hspace*{-0.5cm}
{\bf Fig.10}

Thermally averaged tunneling probability based on
eq.(\ref{finite T penetration}) with $f(t) = 1$ in the text
taking $\omega_* = \omega_B$.
Crossed points are calculated using the approximate formula
(\ref{approximate probability for finite t}).

\vspace{0.5cm} 
\hspace*{-0.5cm}
{\bf Fig.11}

Time averaged dynamical factor for a few models.

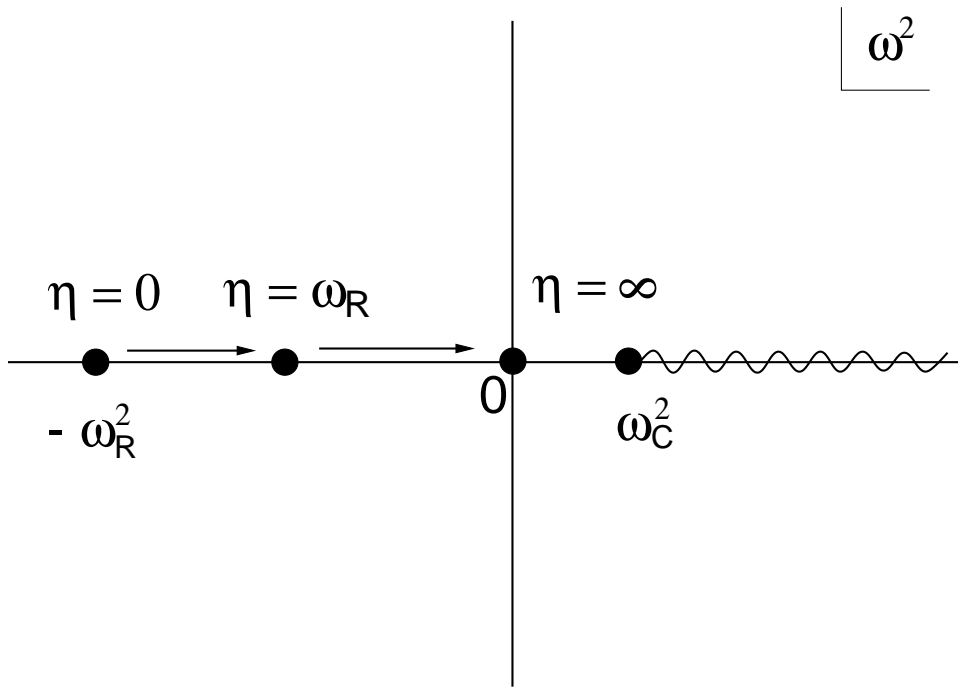
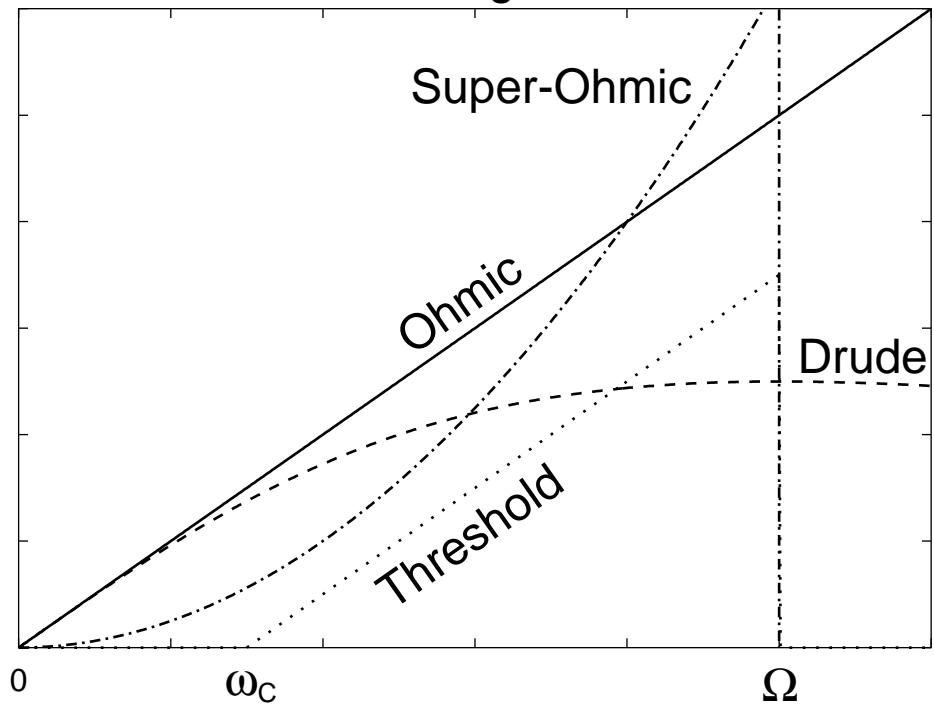
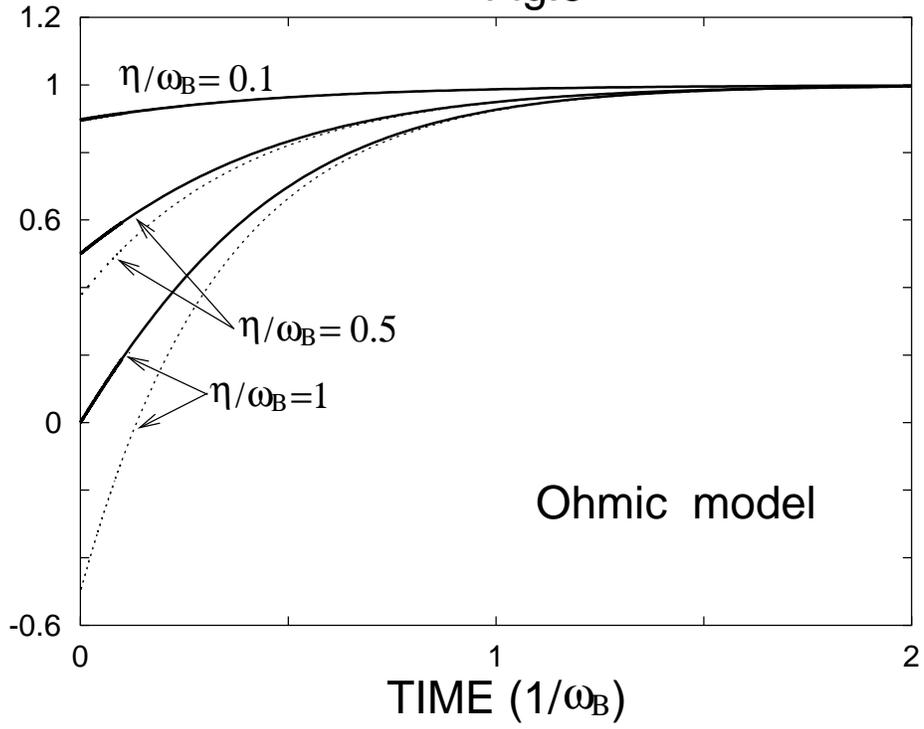
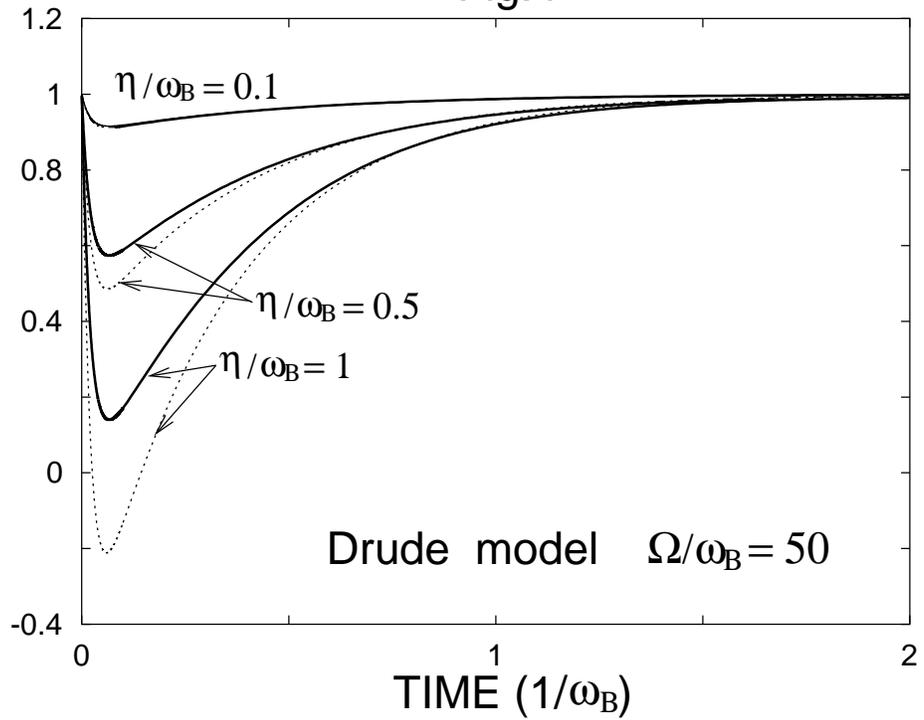
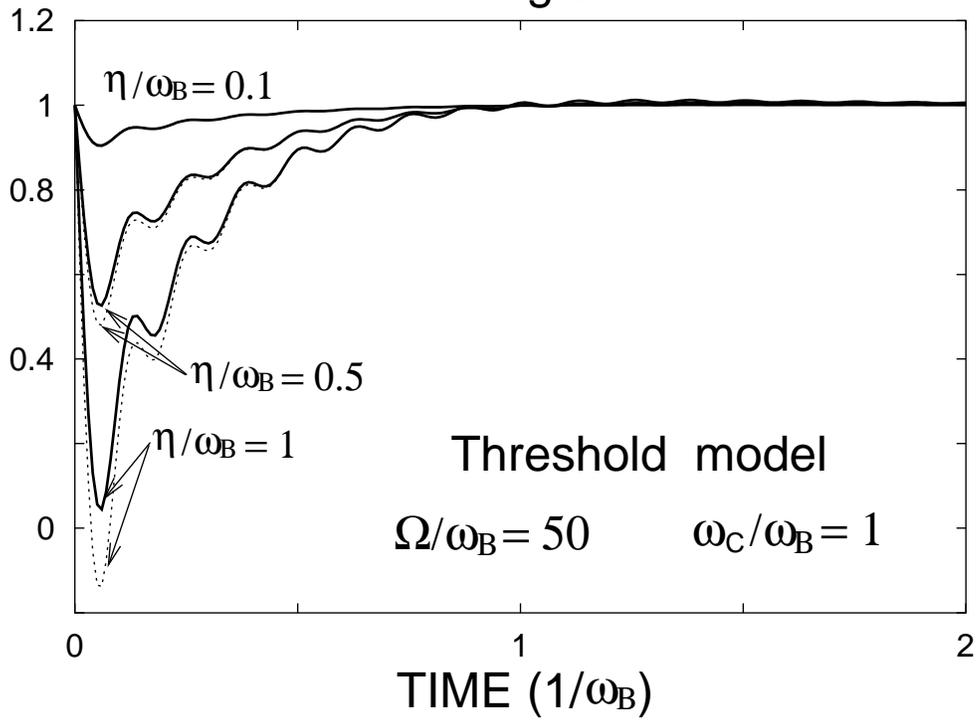
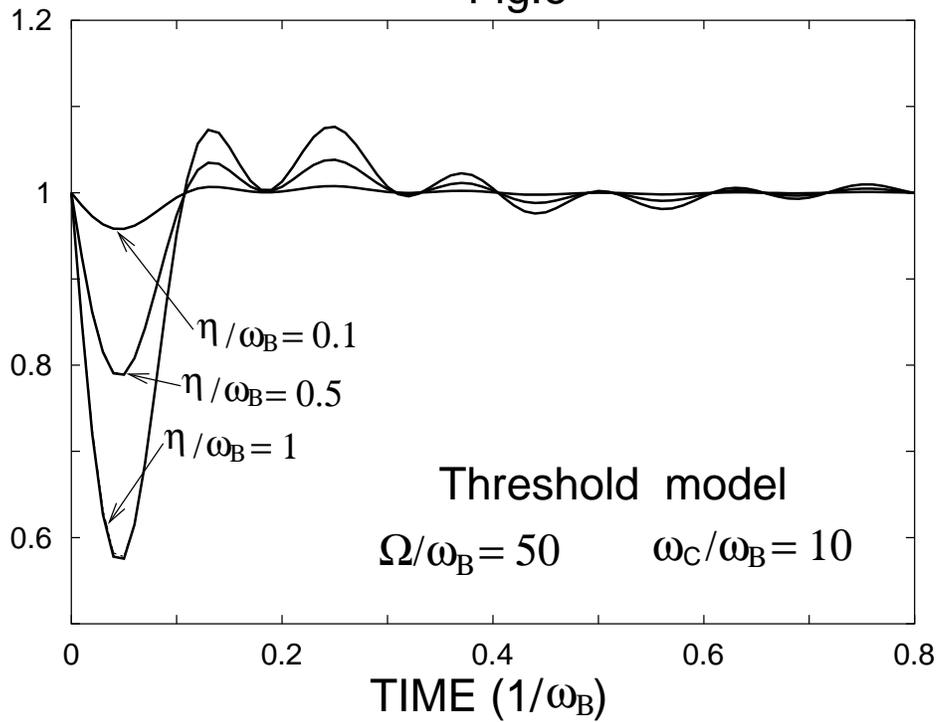
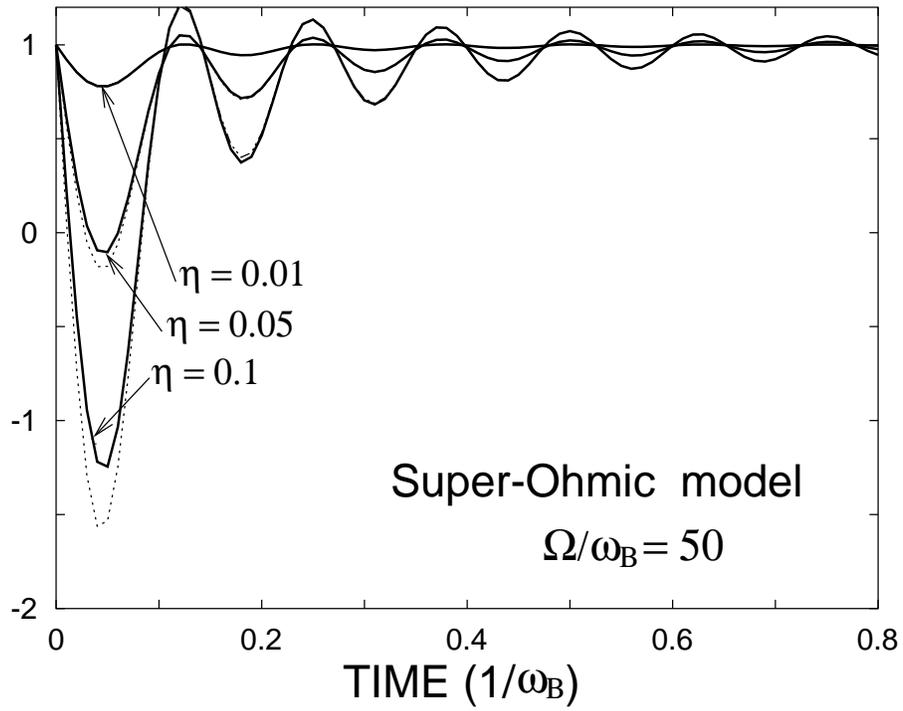
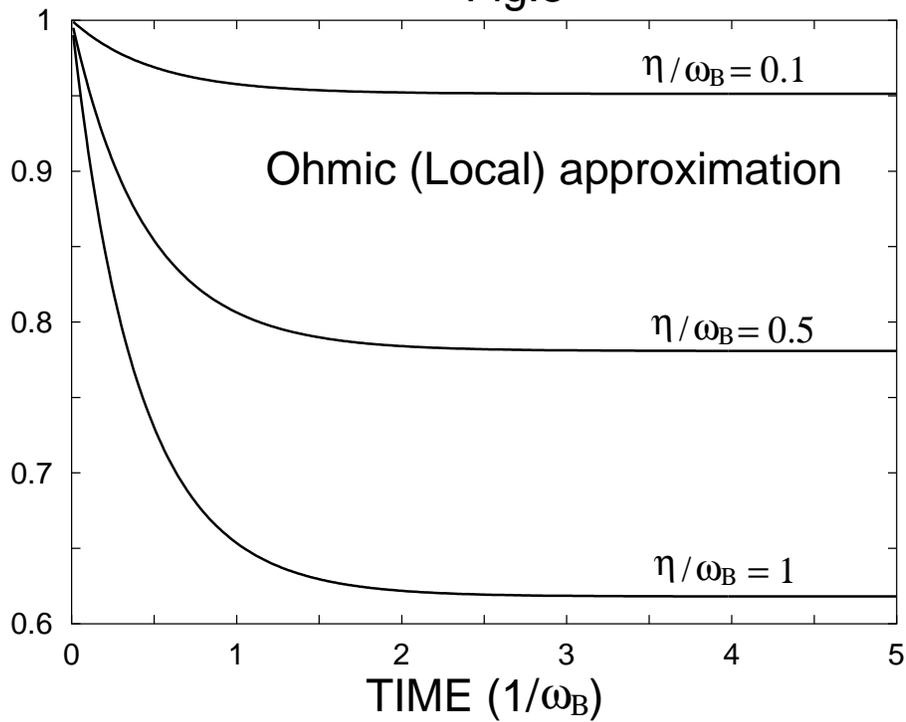
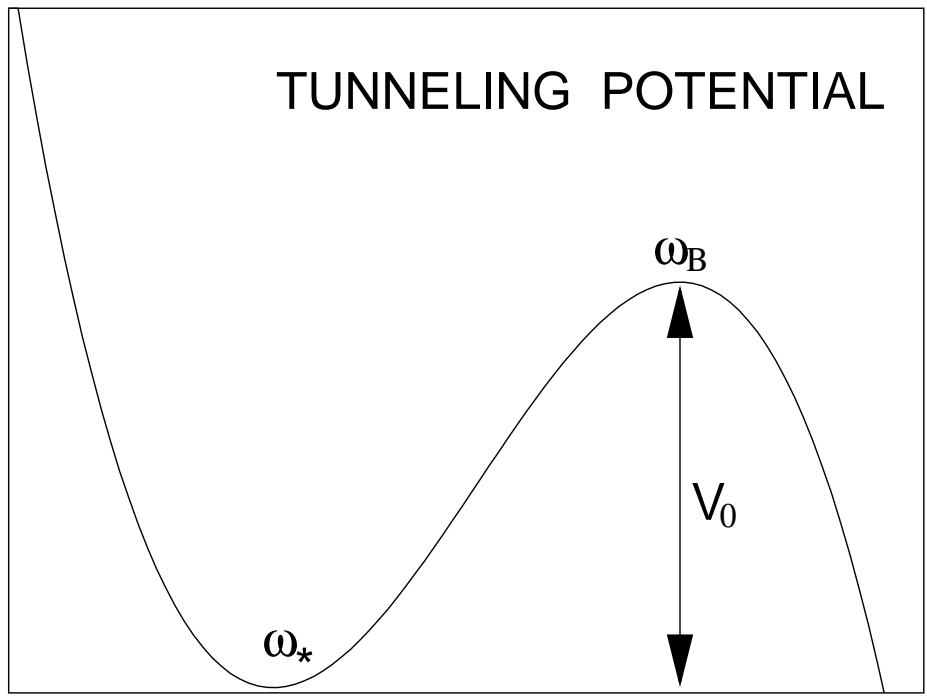
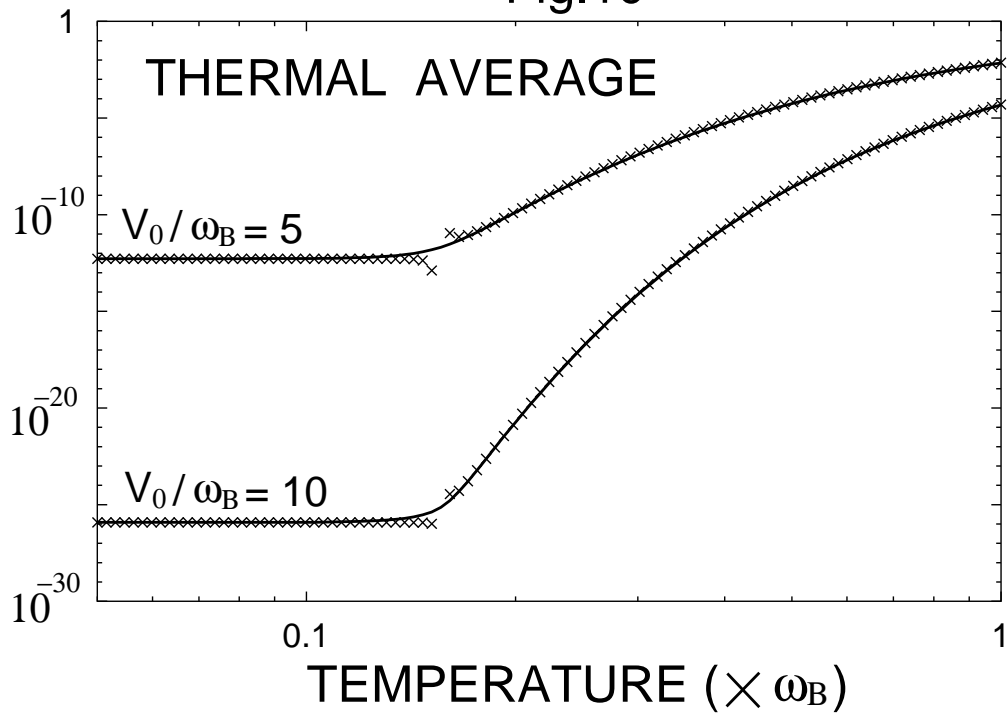
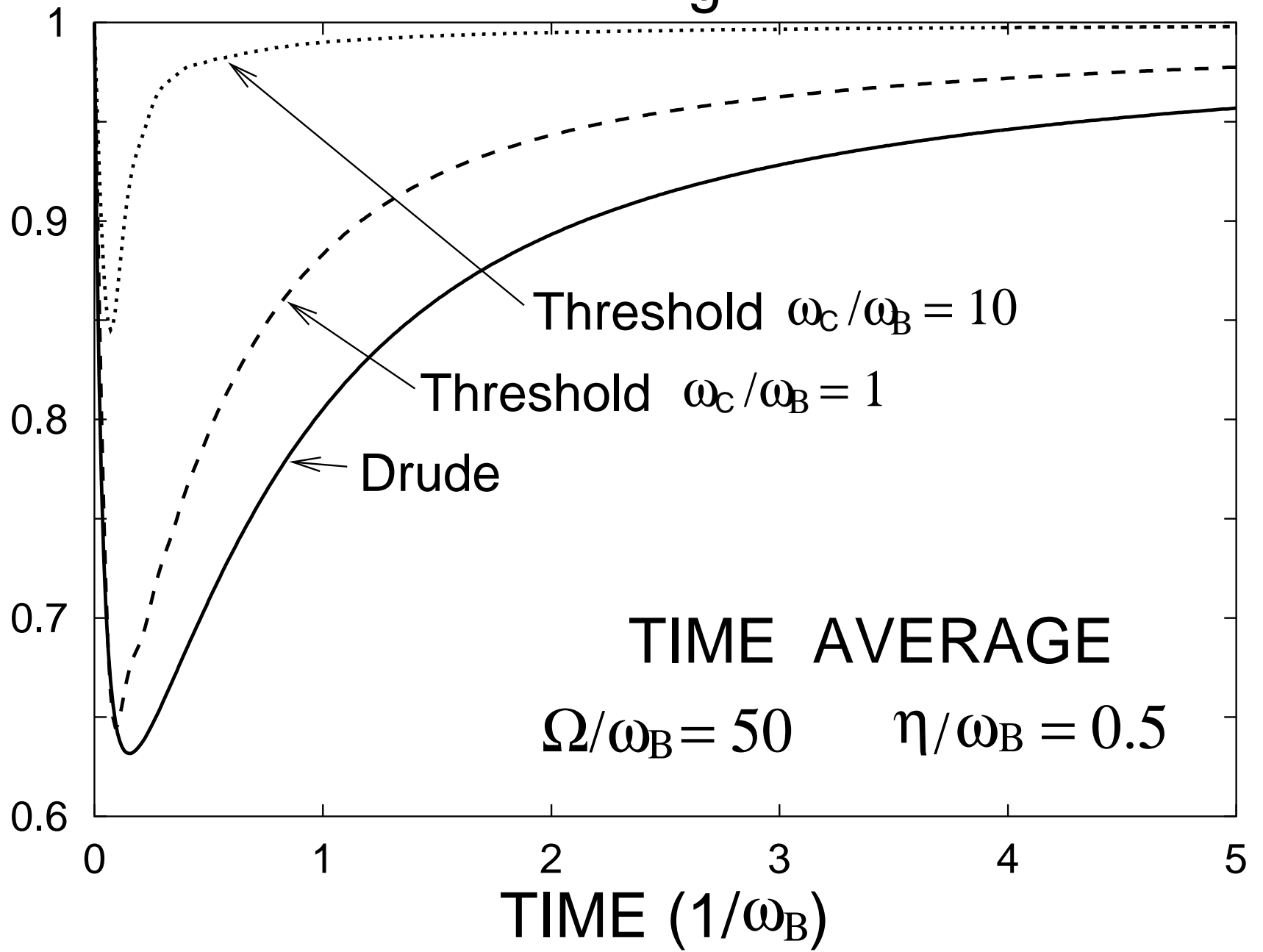
\end{document}